\pdfoutput=1
\documentclass[12pt]{article}
\usepackage{graphicx}
\usepackage{amssymb,amsmath,amsfonts,palatino,amsthm}
\usepackage{amssymb}
\usepackage{epstopdf}
\newcommand{\beq}{\begin{equation}}
\newcommand{\eeq}{\end{equation}}

\newcommand{\f}{\begin{equation}}
\newcommand{\ff}{\end{equation}}

\newtheorem{theorem}{Theorem}
\newtheorem{corollary}{Corollary}
\newtheorem{condition}{Condition}

\begin{document}

%%%%%%%%%%%%%%%%%%%%%%%%%%%%%%%%%%%%%%%%%%%%%%%%
\title{Propagation and interaction of chiral states in quantum gravity}
\author{
Lee Smolin\thanks{Email address: lsmolin@perimeterinstitute.ca} and
Yidun Wan\thanks{Email address: ywan@perimeterinstitute.ca}
\\
\\
\\
Perimeter Institute for Theoretical Physics,\\
31 Caroline st. N., Waterloo, Ontario N2L 2Y5, Canada, and \\
Department of Physics, University of Waterloo,\\
Waterloo, Ontario N2J 2W9, Canada\\}
\date{October 5, 2007}
\maketitle
\vfill
\begin{abstract}
We study the stability, propagation and interactions of braid states
in models of quantum gravity in which the states are four-valent
spin networks embedded in a topological three manifold and the
evolution moves are given by the dual Pachner moves. There are
results for both the framed and unframed case. We study simple
braids made up of two nodes which share three edges, which are
possibly braided and twisted. We find three classes of such braids,
those which both interact and propagate, those that only propagate,
and the majority that do neither.
\end{abstract}
\vfill
\newpage
\tableofcontents
\newpage

\section{Introduction}

There is an old dream that matter is topological excitations of the geometry
of spacetime. Recently it was discovered that this is realized in the context
of models of quantum geometry based on spin networks, such as those used in
loop quantum gravity and spin foam models\cite{fotini,fotini-david,fotsunlee}.
Although the rough idea that topological features of spacetime qeometry
might be interpreted as particles is very old, two questions delayed
implementation of the idea till recently. First, how do we identify an
independent excitation or degree of freedom in a background independent
quantum theory, where the semiclassical approximation is expected to be
unreliable? Second, why should such excitations be chiral, as any excitations
that give rise to the low mass observed fermions must be?

Recently a new approach to the first question was proposed in
\cite{fotini, fotini-david}, which was to apply the notion of
noisefree subsystem from quantum information theory to quantum
gravity, and use it to characterize an emergent elementary particle.
The key point is that noiseless subsystems arise when a splitting of
the hilbert space of the whole system (in this case the whole
quantum geometry) into system and environment reveals emergent
symmetries that can protect subsystems from decohering as a result
of inherently noisy interactions with the environment. The proposal
raised the question of whether there were models of dynamical
quantum geometry in which this procedure reproduced at least some of
the symmetries of the observed elementary particles.

This led to an answer to the second question. In \cite{fotsunlee} it
was shown that this was the case in a simple model of the dynamics
of quantum geometry. The result of that paper incorporated prior
results of \cite{sundance} where a preon model was coded into a game
of braided triplets of ribbons. The needed ribbon graphs are present
in models related to loop quantum gravity when the cosmological
constant is non-zero\cite{qdef,fotlee-dual}; and the three ribbon
braids needed are indeed the simplest systems in that model bound by
the conservation of topological quantum numbers. Strikingly, the
results of \cite{fotsunlee} make contact with both structures from
topological quantum computing and now classic results of knot theory
from the 1980s. In both chirality plays a key role because braids
are chiral and topological invariants associated with braided
ribbons are able to detect chirality and code chiral conservation
laws.

Many questions remained unanswered however. One was quickly answered in
\cite{jonathan1}: can the excitations be considered local in the quantum geometry?
The answer is that many of the braided excitations can be evolved by local
moves to states which can be considered local because they are attached by only a single
edge to the rest of the graph representing the quantum geometry of space. The
same paper answered in the affirmative another question: do these excitations
propagate? If the quantum geometry of space can be modeled as a large and
complicated spin network graph, \cite{jonathan1} showed that the braids which code
the excitations propagate on the larger graph under the local dynamical moves
of the model.

However, the results of \cite{fotsunlee} suffered from a serious
limitation. The conservation laws which preserve the excitations are
exact, which means that there can be no creation and annihilation of
particles. Indeed, as shown by \cite{jonathan1} the braided
excitations of \cite{fotsunlee} are like solitons in integrable
systems: they pass right through each other. This means that
interactions necessary to turn the game coded in \cite{fotsunlee}
into a real theory of the standard model do not appear to exist in
that model.

As a result, a search has been underway for a modification of the dynamics
studied in \cite{fotsunlee} which would allow propagation and interactions of
emergent chiral matter degrees of freedom\cite{JLS-inprogress}. This is not so easy as many
extensions of the local moves studied in \cite{fotsunlee} destabalize the
braids so that there are no longer any emergent conservation laws.

In this paper we report on one successful such modification. This is based on
two ideas proposed by Markopoulou\cite{fot-personal}. First extend the graphs
in the model from three valent to four valent and base the dynamics on the
dual Pachner moves naturally associated with four valent graphs. Second, only
allow the dual Pachner moves when acting on a subgraph which is
dual to a triangulation of a trivial ball in $R^{3}$.

Four-valent graphs and the corresponding dual Pachner moves naturally occur in
spin foam models\cite{spin-foam}, hence this extends the results on the existence of emergent,
chiral degrees of freedom to that much studied case. As we will show below,
the dual Pachner moves, with the restriction to duals of triangulations of
trivial regions, is exactly right to preserve the stability of certain braid
states, while giving some propagation and interactions. We do not however, in
this paper, investigate the correspondence to the preon models of
\cite{sundance,fotsunlee}.

In many studies of spin foam models the graphs and spin foam
histories are taken to be abstract, or unembedded. Our results do
not directly apply to these models, as the topological structures
our results concern arise from the embedding of the graphs in a
topological three manifold. But neither do such models give dynamics
for states of loop quantum gravity, which are embedded. A path
integral formulation of loop quantum gravity must give evolution
amplitudes to embedded graphs, as those are the states found by
quantizing diffeomorphism invariant gauge theories such as general
relativity.

We study here two cases of four valent embedded graphs. The first
are framed, which means that the edges are represented by tubes
which meet at nodes which are punctured spheres. If we take the
limit where the edges become curves, while the nodes remain
punctured spheres we get the unframed case.

One might go further and consider the case in which the nodes are
structureless points rather than punctured spheres. However, in this
case the dual Pachner moves are not all well defined in the embedded
case, so that there cannot be an application of spin foam models to
amplitudes for such graphs which incorporate dual Pachner moves.

The main results of this paper are  as follows:

\begin{itemize}

\item{} The dual Pachner moves on embedded four valent graphs may be restricted to those that are dual to Pachner moves on triangulations of regions of $\mathbb{R}^3$, in such a way that some chiral braid states are locally stable, which means they cannot be undone by local moves involving the nodes that comprise them.

\item{}Some of those stable chiral braids propagate, in the sense that under the local evolution moves they can exchange places with substructures adjacent to them in the graph. The propagation in most cases is chiral. In these cases a braid will propagate along an edge in the larger graph only to the left and its mirror image (which is a distinct state) will propagate only to the right.

\item{}Results are found which limit the classes of propagating states.

\item{}In some cases two braids adjacent to each other in a spin network may merge under the action of the local moves. For this to happen one of them must be in a small class of states called actively interacting. We find examples of actively interacting braids, all of which are equivalent to trivial braids with twists.

\end{itemize}

This is the second of a series of papers in which the conservation,
propagation and interactions of four valent spin networks are
investigated. The first paper \cite{yidun} contains results on
classifications of the relevent braid embeddings and it also
establishes our notation. For completeness we give a summary of the
notation and main results of that paper in the next section. The
dynamics is then introduced in section 3, by giving the allowed set
of dual Pachner moves by which embedded graphs are allowed to
evolve. In section 4 we discuss the local stability of the braids
under the equivalence and evolution moves and introduce a quantity
that is conserved under both. Section 5 gives examples of braids
propagating as well as some results which limit which braids can
propagate. Section 7 does the same with interactions.

\section{Previous results}

We work with a graphic calculus which describes 4-valent framed
spin-networks embedded in 3-manifolds up to diffeomorphisms. The
graphs represent the networks and their embeddings in terms of two
dimensional diagrams which represent projections of the graphs onto
a two-plane. As the spin or, more generally, representation,
representation labels play no role in the results of this paper,
they are not indicated in these diagrams. The notation we use was
described in detail in \cite{yidun}, in this section we review the
notation and the results we will need from that paper.

When using this notation it is important to keep in mind the
distinction between braids, which are diffeomorphism equivalence
classes of embeddings of diagrams  in three dimensional space, and
the braid diagrams that represent them.  Many braid diagrams will
correspond to the same braid, these will be related by a set of
equivalence moves. It will be important also to distinguish these
equivalence moves from dynamical moves we introduce in the next
section, which take equivalence classes, or braids, to equivalence
classes.

\subsection{Notation}

\begin{figure}[h]
\begin{center}
\includegraphics[
natheight=1.434700in,
natwidth=2.948200in,
height=1.4702in,
width=2.9914in
]{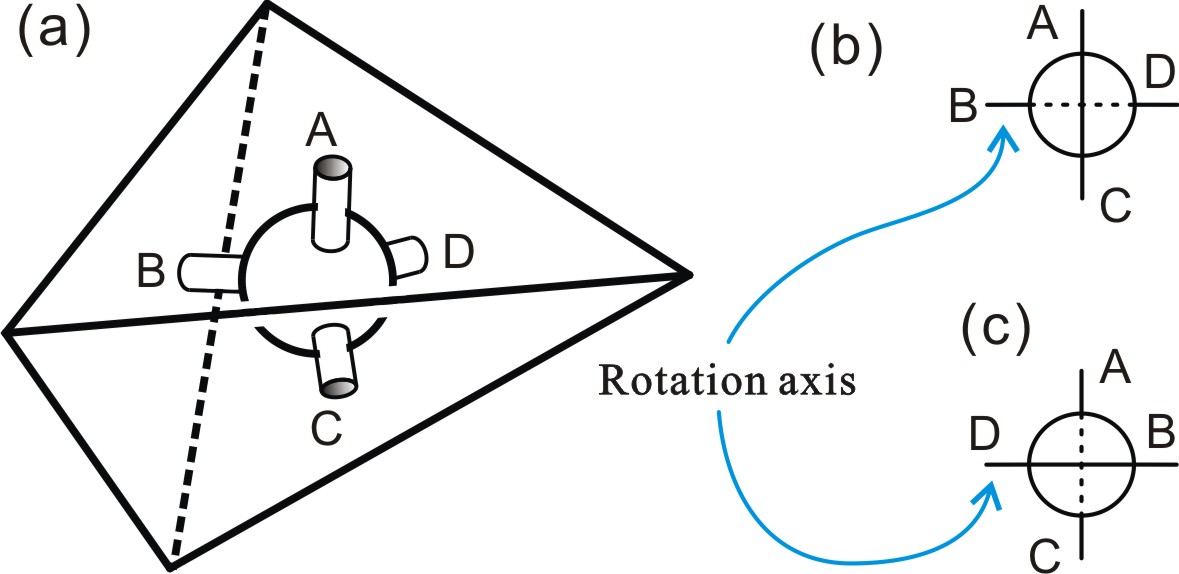}
\end{center}
\caption{(a) is a tetrahedron and its dual node. There are two
orientations of a node that can appear in a diagram (b) denotes
the $\oplus$ state, while (c) denotes the $\ominus$ state.}
\label{notation}
\end{figure}

In the category of framed graphs we discuss edges that are
represented by tubes and nodes that are represented by  2-spheres
with the incident edges attached at circles called punctures. We
consider only four valent nodes which are non-degenerate so that not
more than two edges of a node are co-planar.  Since we are
interested in projections representing graphs up to diffeomorphisms,
there is a single diffeomorphism class of nodes. For convenience,
the nodes are all represented as rigid and dual to tetrahedra, as
shown in Fig. \ref{notation}.  We will limit ourselves to
projections in which all nodes are in one of the configurations or
states shown in Fig. \ref{notation}. Because of the freedom to make
a diffeomorphism before projecting this can be done without loss of
generality. We note that as shown in Fig. \ref{notation} the
different such allowed projections are related to each other by
rotations of $\pi/3$ around an axis defined by one of the edges.

\subsection{Representation of twists}

\begin{figure}[h]
\begin{center}
\includegraphics[
natheight=1.577400in,
natwidth=2.945600in,
height=1.6137in,
width=2.9888in
]{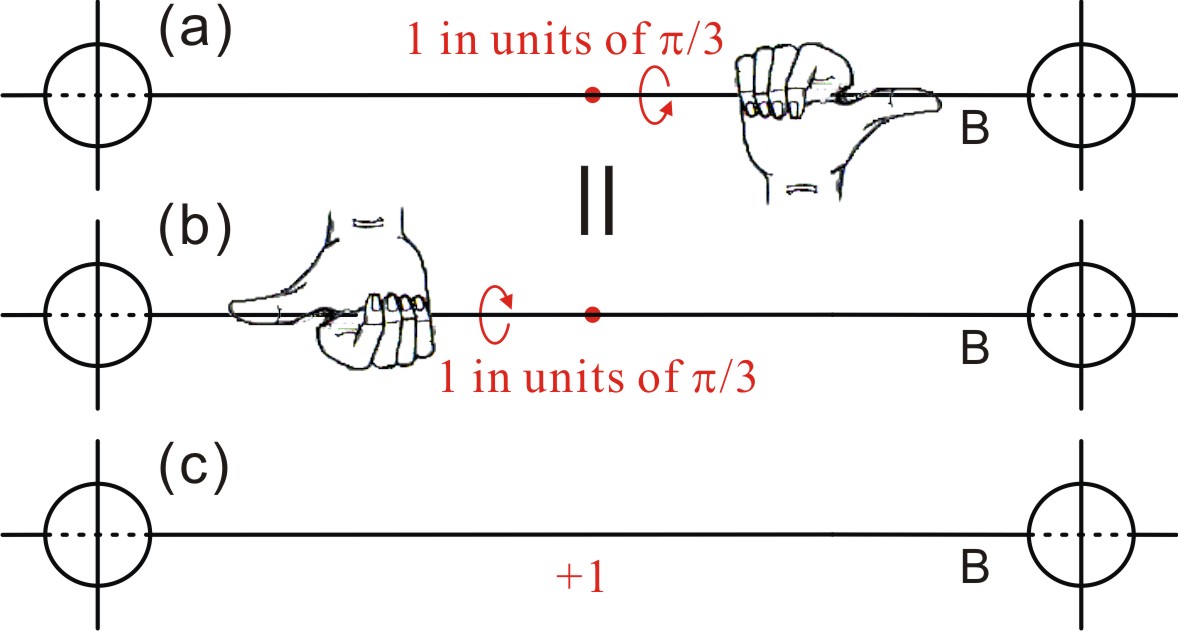}
\end{center}
\caption{The 1 unit of twist (equivalent to a $\pi/3$ rotation) in
(a) means cut to the right of the red dot and twist as shown. This
is is equivalent to the opposite twist on the opposite side of the
red dot, as shown in (b). Thus, both may be represented as in (c),
by a label of $+1$ of edge $B$.}
\label{notationtwist}
\end{figure}

As we have noted, the possible states by which a node may be
represented in a projection can be taken into each other by $\pi/3$
rotations around one of the edges. By the local duality of nodes to
tetrahedra, these correspond to the $\pi/3$ rotations that relate
the different ways that two tetrahedra may be glued together on a
triangular face. These rotations cause twists in the edges and, as a
result of the restriction on projections of nodes we impose, the
twists in a projection of an edge of a graph will be in units of
$\pi/3$. This allows a simple representation of twists of edges
which is  shown in Fig. \ref{notationtwist}(a).

We note that the twist indicated in \ref{notationtwist}(a) is
equivalent to that shown in \ref{notationtwist}(b). This provides a
way simplifying the notation, as it allows us to label an edge with
a right-handed twist by a positive integer (and left-handed twist by
a negative integer). Thus, Fig. \ref{notationtwist}(a) and (b) can
be replaced by \ref{notationtwist}(c) without introducing any
ambiguity.

\subsection{Framed and unframed graphs}

Most of the results of this paper will refer to the case of framed
graphs, defined above. However, unframed graphs are used in loop
quantum gravity and it is useful to have results then for that case
as well. To do that we need a particular definition of unframed
graphs, which is gotten from the framed case discussed here by
dropping information about rotations or twists of the edges, but
keeping the nodes as spheres, locally dual to tetrahedra. This is
necessary so that the dual Pachner moves are well defined for
embedded graphs. If we drop the resolution of nodes as spheres, and
represent them as structureless points which are just intersections
of edges, we loose the information as to which pairs of edges at a
node are over or under (in the sense of Fig. \ref{notation}) and the
dual Pachner moves which will be defined in the next section will no
longer well defined.

In the rest of this paper we refer always to the framed case, unless we
explicitly describe results for the unframed case.

\subsection{Braids}

\begin{figure}[h]
\begin{center}
\includegraphics[
natheight=0.909800in,
natwidth=2.764800in,
height=0.9426in,
width=2.808in
]{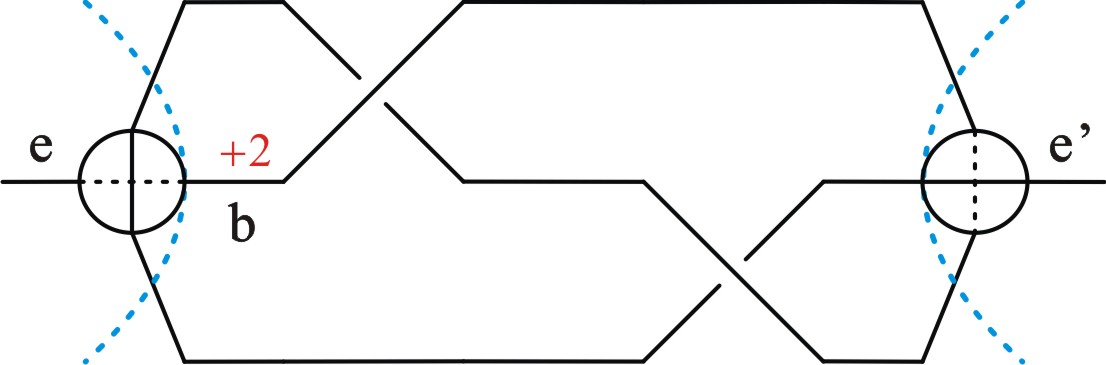}
\end{center}
\caption{A typical 3-strand braid diagram formed by the three common
edges of two end-nodes. The region between the two dashed line
satisfies the definition of an ordinary braid. Edges $e$ and
$e^{\prime}$ are called external edges. There is also a right handed
twist of 2 units on strand $b$. In this figure the left handed node
is in an $\oplus$ state while the right handed node is in an
$\ominus$ state.} \label{braid}
\end{figure}

We will be interested in a class of embedded sub-spinnets which
consist of two nodes which share three edges which may be braided.
These are defined as braids, whose two dimensional projections are
called braid diagrams\cite{yidun}. An example of such a braid
diagram is shown in Fig. \ref{braid}. In the rest of the paper, we
use braids for both braids and their braid diagrams, unless an
emphasis on braid diagrams is necessary.

\subsection{Crossings and chirality}

\begin{figure}[h]
\begin{center}
\includegraphics[
natheight=0.460100in,
natwidth=2.940400in,
height=0.4903in,
width=2.9836in
]{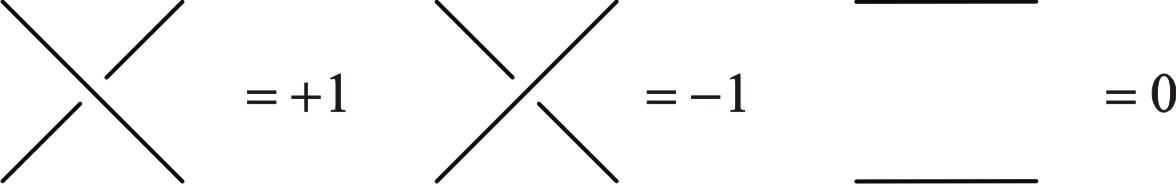}
\end{center}
\caption{The assignments of right-handed crossing, left-handed crossing, and
null crossing respectively from left to right.}%
\label{assignment}%
\end{figure}

As usual we can assign a number to a crossing according to its chirality, viz
$+1$ for a right-handed crossing, $-1$ for a left-handed crossing, and $0 $
otherwise. Fig. \ref{assignment}\ shows this assignment.

\subsection{Equivalence moves}

Diffeomorphism classes of embedded knots and links are described in
terms of two dimensional diagrams which represent projections of the
embedded graphs onto a two dimensional plane. Many projections,
hence many diagrams, represent the same diffeomorphism equivalence
class of an embedded grap. In the case of knots and links without
nodes this is reflected in Reidemeister's theorem which gives a
small number of local moves in the diagrams such that any two
diagrams represent the same three dimensional diffeomorphism
equivalence class iff they are related by a finite sequence of such
moves.

For the class of framed four valent graphs with rigid nodes, the Reidemeister
moves are extended by two classes of moves\cite{yidun}.

\begin{figure}[h]
\begin{center}
\includegraphics[
natheight=1.241900in,
natwidth=2.952500in,
height=1.2773in,
width=2.9966in
]{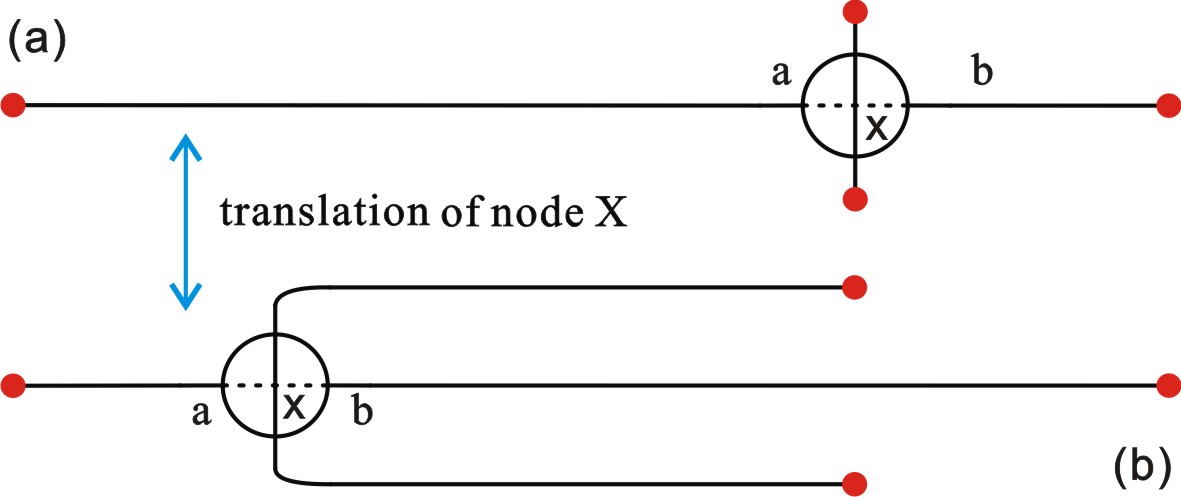}
\end{center}
\caption{Red points I and J represent other nodes where edges $a$ and $b$ are
attached to. (b) is obtained from (a)\ by translating node $X$ from right to
left, and vice versa.}%
\label{translation}%
\end{figure}

\begin{figure}[h]
\begin{center}
\includegraphics[
natheight=0.952200in,
natwidth=2.948200in,
height=0.985in,
width=2.9914in
]{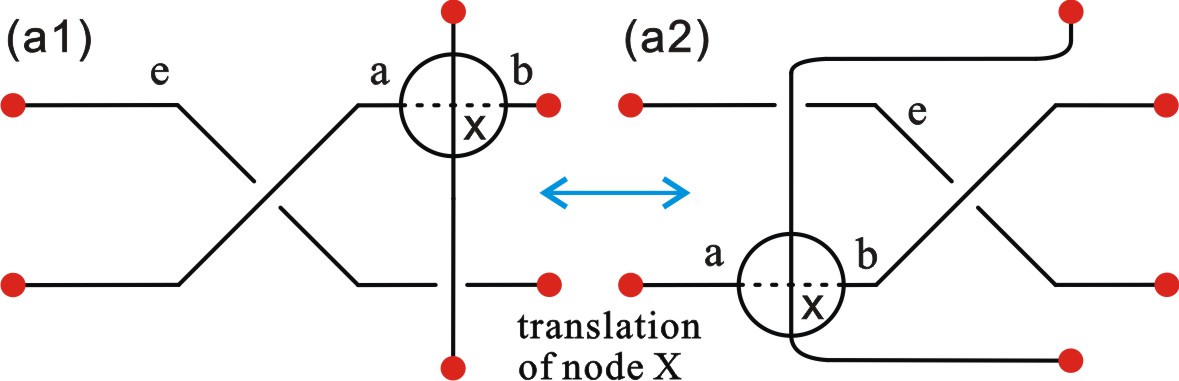}
\end{center}
\caption{Red points represent other nodes where edges $a$ and $b$ are attached
to. (a1) and (a2) can be transformed into each other by translating node $X$.}%
\label{translationX}%
\end{figure}

\begin{figure}[h]
\begin{center}
\includegraphics[
natheight=2.082500in,
natwidth=2.943000in,
height=2.1101in,
width=3.5699in
]{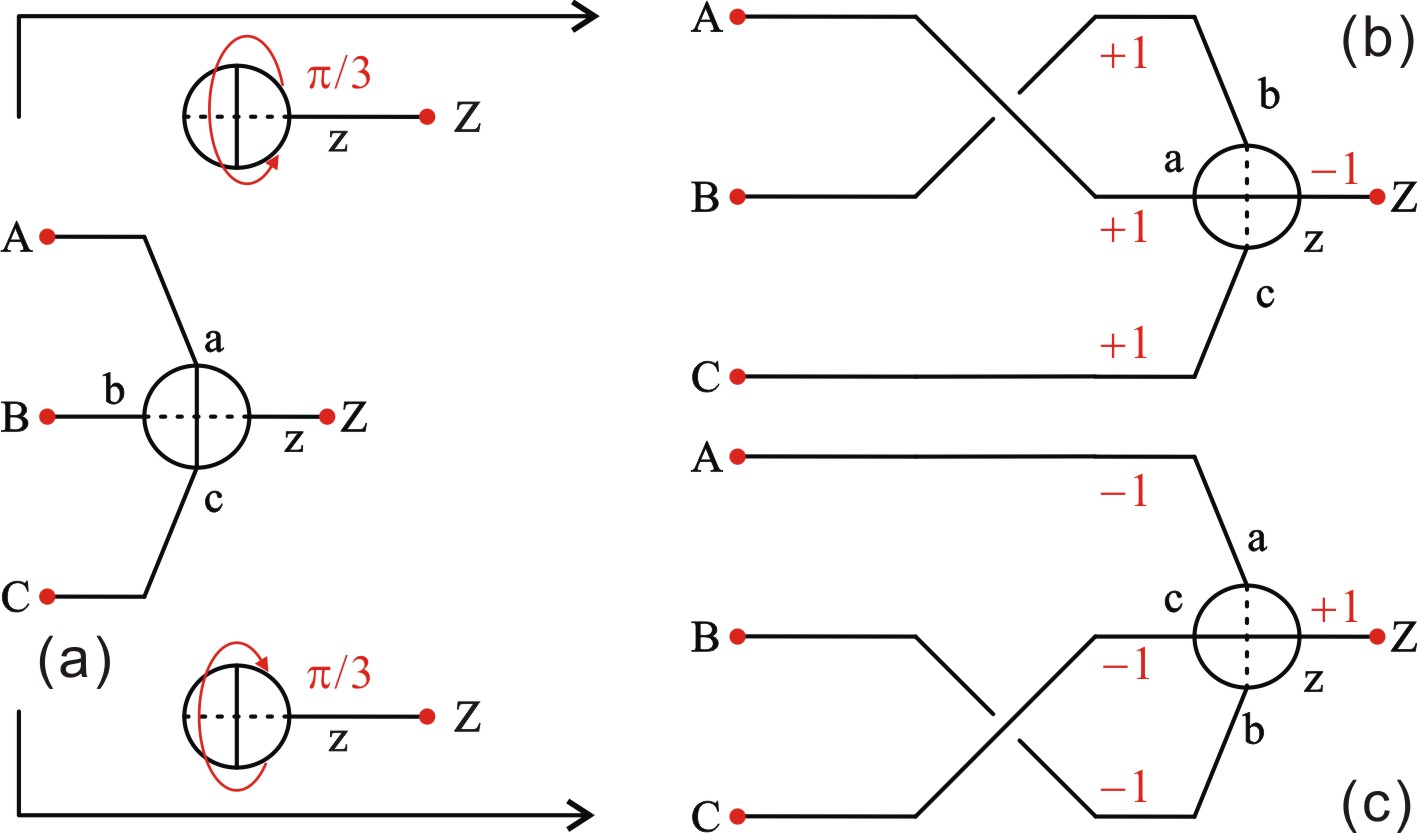}
\end{center}
\caption{(b) \& (c) are results of (a) by rotating the $\oplus$-node in (a)
w.r.t. edge $z$ in two directions respectively. Points $A $, $B$, $C$, and $Z
$ are assumed to be connected somewhere else and are kept fixed during the
rotation. All edges of the node gain the same amount of twist after rotation.
Note that a $\pi$-rotation changes the state of a node, i.e. if a node is in
state $\oplus$ before the rotation, it becomes a $\ominus$-node after the
rotation.}%
\label{pi3rot+}%
\end{figure}

\begin{figure}[h]
\begin{center}
\includegraphics[
natheight=2.147300in,
natwidth=2.943000in,
height=2.1223in,
width=3.5699in
]{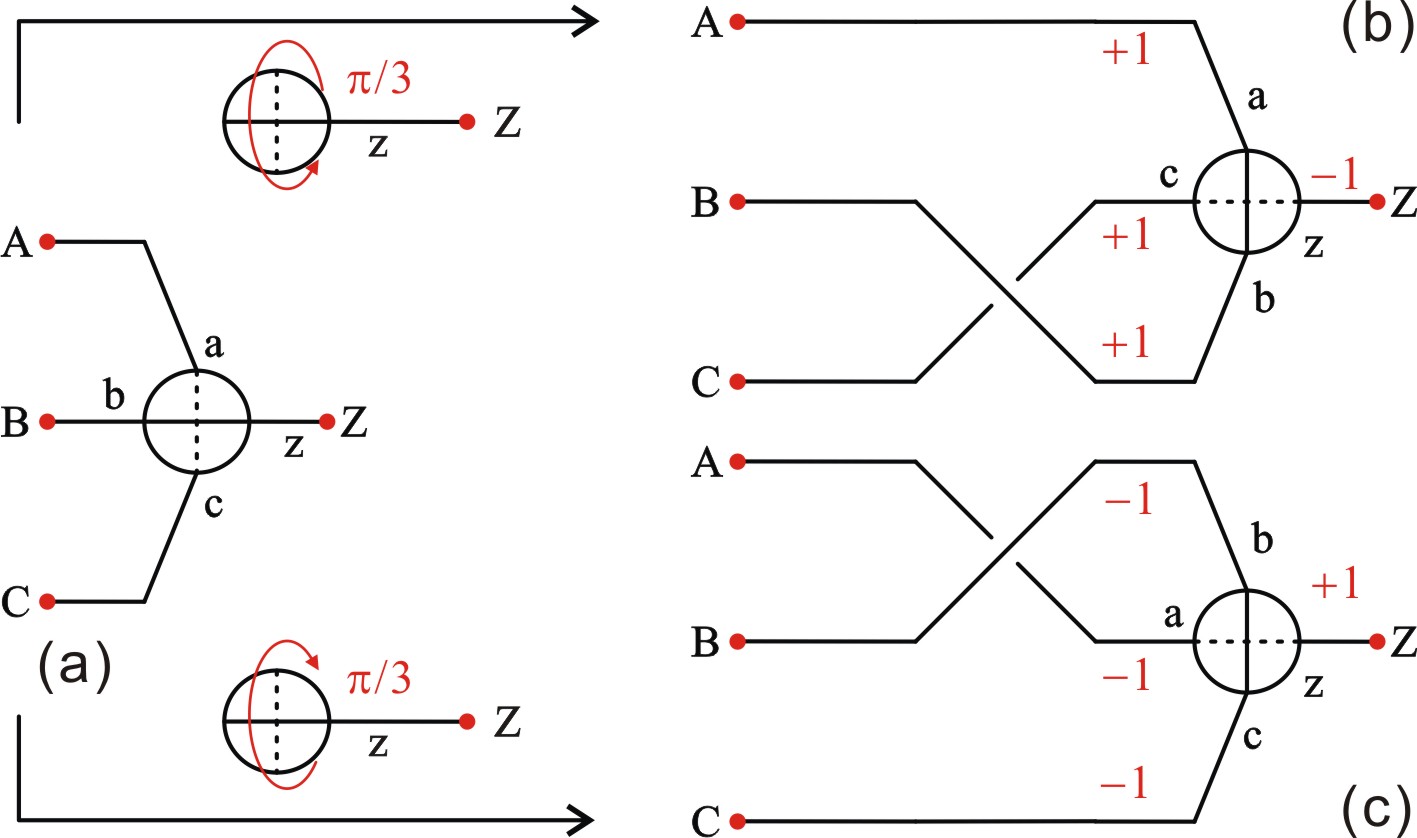}
\end{center}
\caption{(b) \& (c) are results of (a) by rotating the $\ominus$-node in (a)
w.r.t. edge $z$ in two directions respectively. Points $A$, $B$, $C$, and $Z$
are assumed to be connected somewhere else and are kept fixed during the
rotation. All edges of the node gain the same amount of twist after rotation.
}%
\label{pi3rot-}%
\end{figure}

\begin{figure}[h]
\begin{center}
\includegraphics[
natheight=2.092800in,
natwidth=2.932600in,
height=2.1318in,
width=2.9767in
]{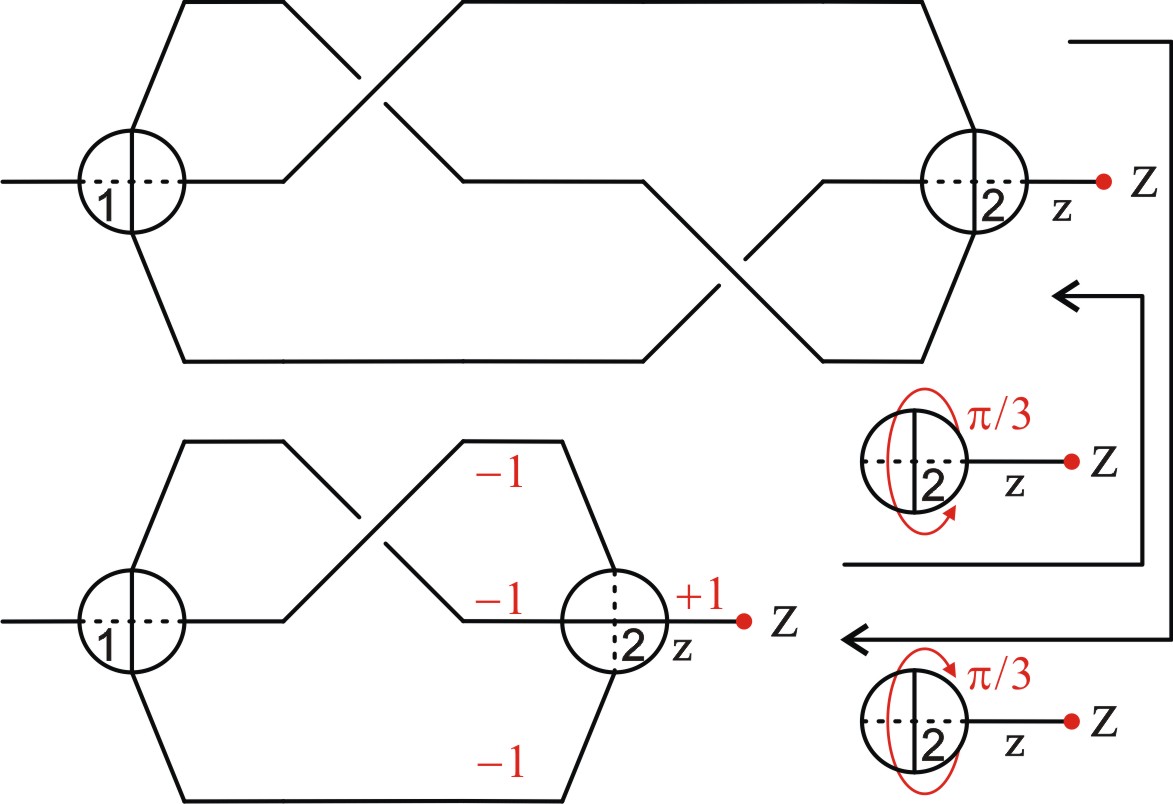}
\end{center}
\caption{The two braid diagrams are equivalent because they can can
be transformed into each other by a $\pi/3$-rotation of node 2.}%
\label{equibraids}%
\end{figure}

\begin{itemize}
\item {}Translations in which nodes are slid over or under crossings in the
diagram, as shown in Figures \ref{translation} and
\ref{translationX}.

\item {}Rotations, which are exhibited in Figures \ref{pi3rot+} and
\ref{pi3rot-} generate equivalences between diagrams as shown in
Fig. \ref{equibraids}.
\end{itemize}

\subsection{Classification of Braids}

A  braid diagrams can be equivalent under
the equivalence moves to a diagram with a different number of crossings. It is
often useful to work with a representative of an equivalence class which has
the fewest number of crossings, this motivates the following definition of irreducible and reducible braids:

\begin{quotation}
\label{defRedubraid}A braid diagram is \textbf{reducible} if it is
equivalent to a braid diagram with a fewer number of crossings;
otherwise, it is \textbf{irreducible}.
\end{quotation}

The braid on top part of Fig. \ref{equibraids} is an example of a reducible
braid, while the  bottom of the figure shows an irreducible braid.

It is convenient to talk about end-nodes of braids, which are a
portion of a braid including one of the two end nodes.  An
$N$-crossing end-node is said to be a \textbf{reducible end-node},
if it is equivalent to an $M$-crossing end-node with $M<N$, by
equivalence moves done on its node; otherwise, it is irreducible.

A braid is \textbf{reducible} if it has a reducible end-node. If a
braid has a reducible left end-node it is called
\textbf{left-reducible}. If it has both a left and a right reducibe
end node it is  \textbf{\ two-way-reducible.}

If a braid  $B$ can be reduced to an unbraid, i.e. a braids with no crossing, $B$ is
said to be \textbf{completely reducible.}

The following theorem\cite{yidun} states that there is no need to investigate
end-nodes with more crossings to see if they are irreducible.

\begin{theorem}
\label{theoNirred}An $N$-crossing end-node, $N>2$, which has an
irreducible 2-crossing sub end-node, is irreducible.
\end{theorem}
\bigskip

\section{Dynamics and evolution moves}

To define the models our results apply to we have to choose a set of
dynamical evolution moves. In spin-foams and other models of
dynamics of spin networks it is common to pick the dual Pachner
moves\cite{spin-foam}. To motivate the form of them we posit here it is useful to
recall how the Pachner moves arise in combinatorial topology. One
defines the topology of a three manifold through operations on a
simplicial triangulation. Many different simplicial complexes
correspond to the same topological three manifold. Given two of
simplicial complexes one wants to know whether they correspond to
the same topological three manifold. Pachner's theorem gives the
answer, it says they do if the two simplicial complexes can be
connected by a finite sequence of local moves, which are called the
Pachner moves. They are illustrated in Figures \ref{2to3tet} and
\ref{1to4+}(a).

The dual to a simiplical complex is a framed four valent graph in which nodes
are dual to tetrahedra and framed edges are dual to faces. One needs the
framing of the graphs to preserve orientations. An edge between two nodes
tells us that two tetrahedra, dual to those two nodes, share a triangular
face. The framing is needed to tell us how to identify the two triangles.

Let us fix a non-singular differentiable manifold $\mathcal{M}$ and
choose a triangulation of it in terms of tetrahedra embedded in
$\mathcal{M}$ whose union is $\mathcal{M}$. Any such simplical
triangulation of $\mathcal{M}$ has a natural dual which is a framed
four valent graph embedded in that manifold, $\mathcal{M}$. If one
makes a Pachner move on the triangulation, that results in a local
move in the framed graph. These are the dual Pachner moves.

However, not every embedding of a framed four valent graph in $\mathcal{M}$ is
dual to a triangulation of $\mathcal{M}$. Examples of obstructions to finding
the dual include the case of two nodes which share three edges, which are
braided, such as shown in Figure \ref{braid}. This is an embedding of a graph that could not have arisin from
taking the dual of a regular simplicial traingulation of $\mathcal{M}$. We
note that these obstructions are local, in the sense that a subgraph of the
embedded graph could be cut out and replaced by another subgraph that would
allow the duality to a triangulation of $\mathcal{M}$.

This leads to a problem, which is how one defines the dual Pachner moves on
subgraphs of embedded graphs which are not dual to any triangulation of
$\mathcal{M}$. The answer we take is that we do not. This leads to the basic rule:

\textit{Basic rule:} The evolution rules on embedded framed four valent graphs
are the dual Pachner moves and they are allowed only on subgraphs which are
dual to a ball in $R^{3}$.

We now discuss in detail the allowed moves.

\subsection{The $2\rightarrow3$ move \label{sec2to3}}

\begin{figure}[h]
\begin{center}
\includegraphics[
natheight=2.792500in,
natwidth=2.960300in,
height=2.8349in,
width=3.0035in
]{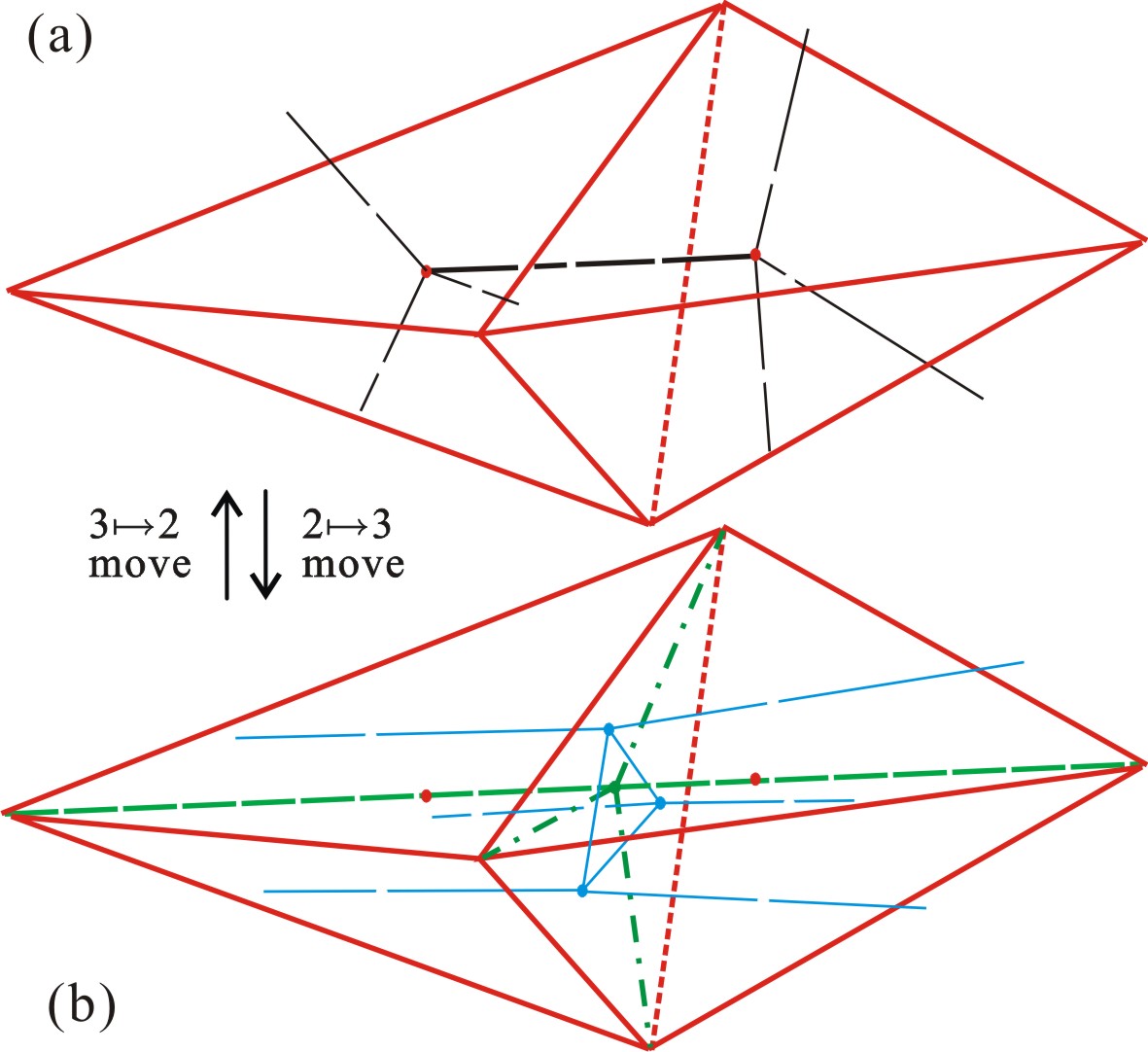}
\end{center}
\caption{(a) shows two tetrahedra with a common face; the black lines represent
the dual graph. (b) shows the three tetrahedra which result from a $2\rightarrow
3$ Pachner move applied to the pair of tetrahedra in (a); the blue lines indicate the dual graph.}
\label{2to3tet}
\end{figure}

In Fig. \ref{2to3tet} we depict the $2\leftrightarrow3$ Pachner move
on tetrahedra and the dual move on four valent graphs. One can see
that as the result of the move, the 2 vetices together with the edge
between in (a) are replaced by three new vertices and three new
edges between them in (b), and vice versa.

The result of a $2\rightarrow3$ move on tetrahedra is unique up to
equivalence moves (e.g. rotations). As a consequence the dual
$2\rightarrow3$ move on framed graph embeddings is unique.

\begin{figure}[h]
\begin{center}
\includegraphics[
natheight=2.264900in,
natwidth=2.945600in,
height=2.3047in,
width=2.9888in
]{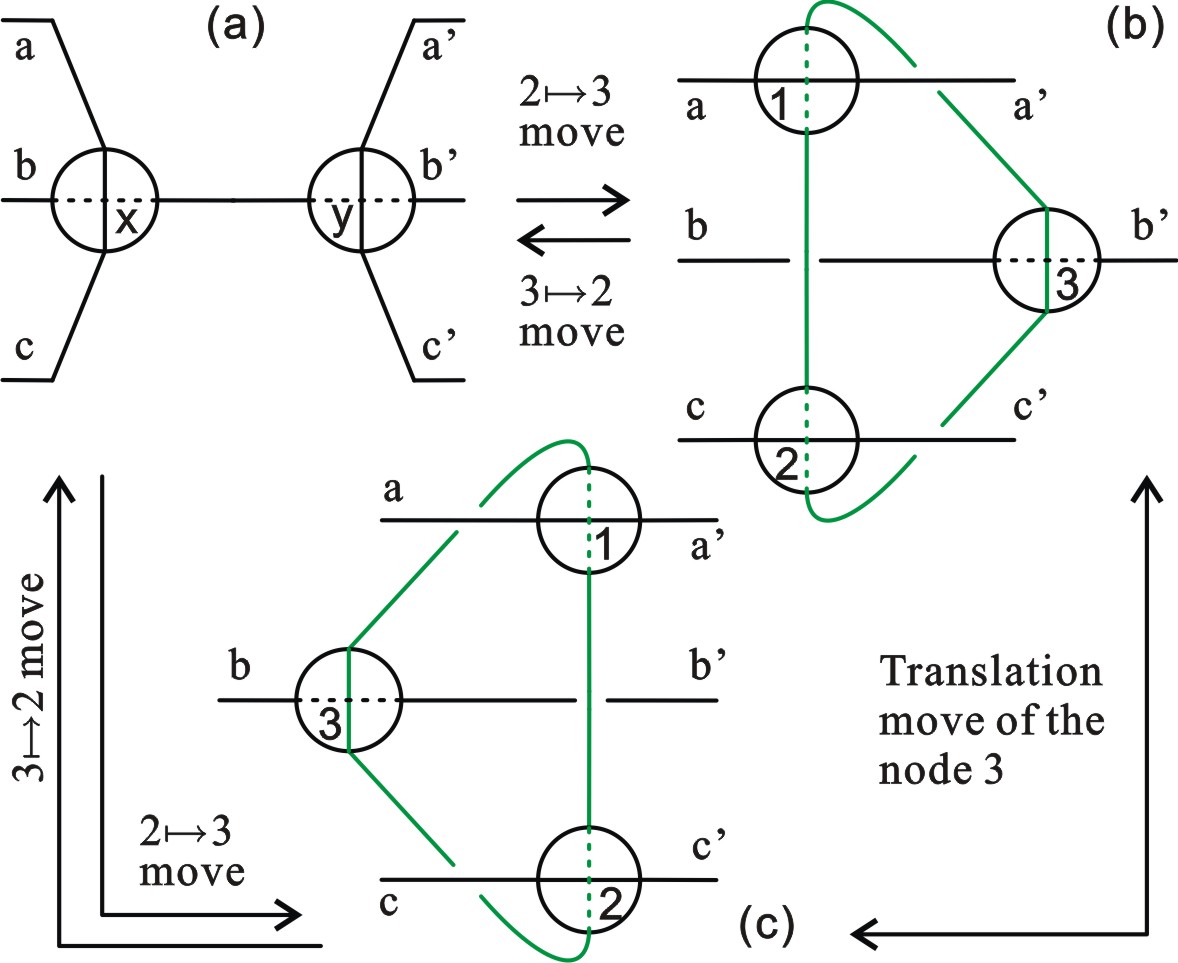}
\end{center}
\caption{(a) is the original configuration of two nodes in
$\oplus$-state, ready for a $2\rightarrow3$ move. (b) is a result of
the $2\rightarrow3$ move from (a); it also leads to (a) via a
$3\rightarrow2$ move. The configuration of nodes $X$ and $Y$ in Fig.
\ref{2to3++}(a) is dual to the two tetrahedra with a common face in
Fig. \ref{2to3tet}(a). Fig. \ref{2to3tet}(b) shows how the two
tetrahedra in (a) are decomposed into three tetrahedra, two of which
are in the front and one is behind, by the dashed green lines; such
an operation is the $2\rightarrow3$ Pachner move. To represent Fig.
\ref{2to3tet}(b) in its dual picture, i.e. our notation of the
framed spinnets, naturally leads to Fig. \ref{2to3++}(b), in which a
contractible loop is generated (see the green edges). One can also
translate the node 3 in Fig. \ref{2to3++}(b) to the left of the
vertical green edge, which gives rise to Fig. \ref{2to3++}(c). We
can go back to Fig. \ref{2to3++}(a) from (b) and (c) by a reverse
move, viz the $3\rightarrow2$ move. One may notice that in Fig.
\ref{2to3++}(a), the nodes $X$ and $Y$ are all in state $\oplus$ if
the common edge of $X$ and $Y$ is chosen as the rotation axis. We
can also imagine the situation where the two nodes ready for a
$2\rightarrow3$ move, like $X$ and $Y$ in Fig. \ref{2to3++}(a), are
in $\ominus$-state with respect to their common edge; the
corresponding $2\rightarrow3$ and its reverse are shown in
Fig. \ref{2to3--}.}%
\label{2to3++}%
\end{figure}

\begin{figure}[h]
\begin{center}
\includegraphics[
natheight=2.405000in,
natwidth=2.945600in,
height=2.4465in,
width=2.9888in
]
{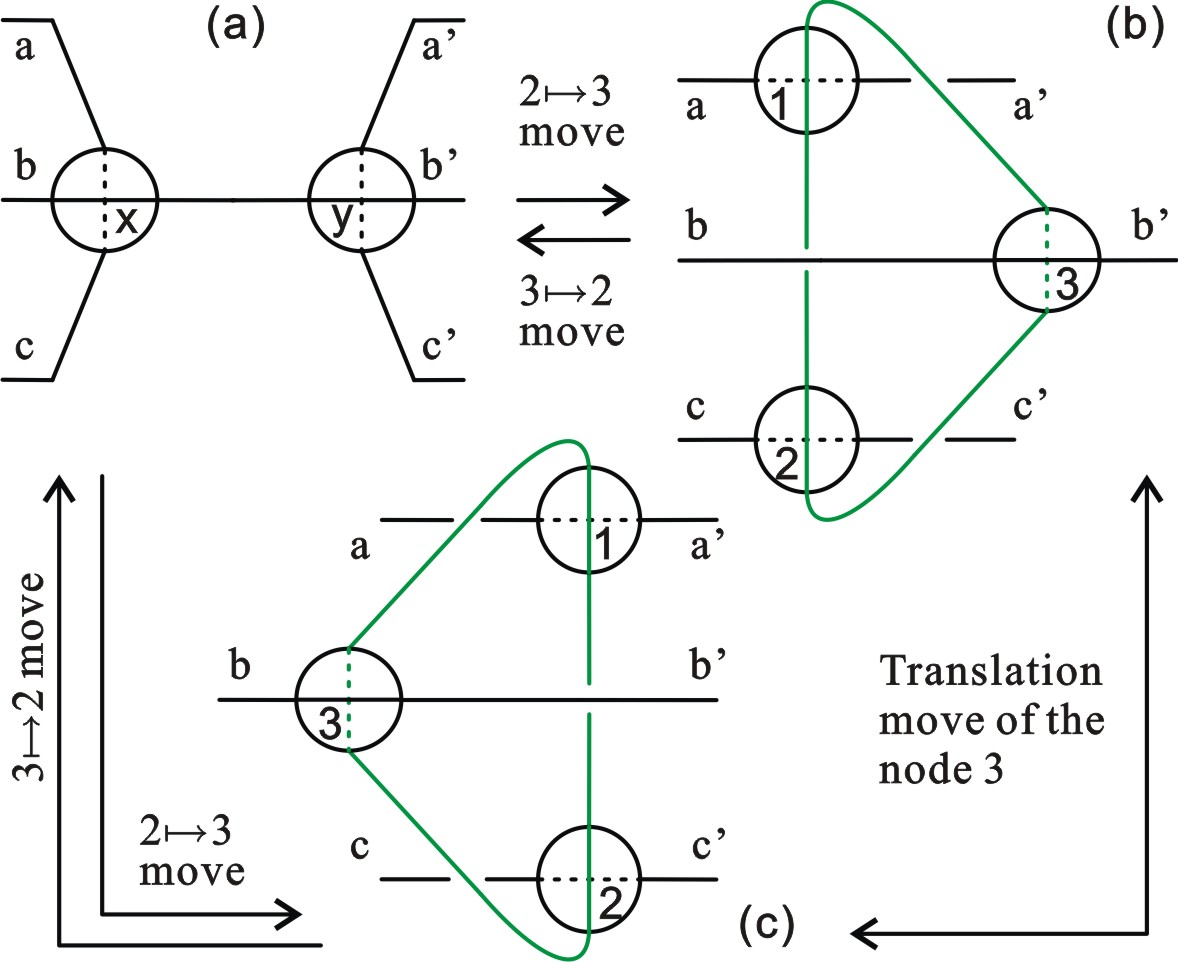}
\end{center}
\caption{(a) is the original configuration of two nodes in $\ominus$-state,
ready for a $2\rightarrow3$ move. (b) is a result of the $2\rightarrow3$ move
from (a); it also leads to (a) via a $3\rightarrow2$ move. }%
\label{2to3--}%
\end{figure}

We next have to translate the allowed $2 \rightarrow3$ dual move to the
diagramatic notation introduced in \cite{yidun} and reviewed in the last
section. This is done in Figures \ref{2to3++} and \ref{2to3--}.

According to the basic rule, a $2\rightarrow3$ move is only doable on two
neighboring nodes with one or more common edges in the case that that subgraph
is dual to a triangulation of a ball in $R^{3}$. In terms of our diagrammatic
notation this translates into the following conditions.

\begin{condition}
\label{con2to3}A $2\rightarrow3$ move is doable on two nodes if and
only if

\begin{enumerate}
\item the two nodes have one and only one common edge and can be arranged in
either of the forms in Fig. \ref{2to3++}(a) or Fig. \ref{2to3--}(a);

\item there is no twist (in the framed case) on the common edge (note that if
there is a twist on the common edge, one has to rotate either of the node to
annihilate the twist);

\item the states of the two nodes with respect to the common edge are either
both $\oplus$ or both $\ominus$.
\end{enumerate}
\end{condition}

The reasoning for the above conditions are easily understood in terms of the
local dual picture of tetrahedron. If Condition \ref{con2to3} holds for two
neighboring nodes, there will be no twist on the edges forming the loop
generated by the $2\rightarrow3$ move, as shown in above figures.

\subsection{The $3\rightarrow2$ move.}

The $3\rightarrow2$ Pachner move is the inverse of the $2\rightarrow3$ Pachner
move. It requires three tetrahedra each pair of which share a face. By taking
the dual one finds a $3\rightarrow2$ move on embedded framed four valent
graphs, it operates on three nodes that are connected in pairs to make a
triangle. The basic rule imposes conditions of a legal $3\rightarrow2$ move
which are as follows.

\begin{condition}
\label{con3to2}A $3\rightarrow2$ move is doable on three neighboring
nodes if and only if

\begin{enumerate}
\item the three nodes with their edges can be arranged in one of the four
proper configurations, namely Fig. \ref{2to3++}(b), (c) and Fig.
\ref{2to3--}(b), (c);

\item the loop formed by common edges of the three nodes is contractible, i.e.
there is no other edges going through the loop or tangled with the
loop. See Figures. \ref{3to2not}(a) \& (b);

\begin{figure}[h]
\begin{center}
\includegraphics[
natheight=1.307600in, natwidth=2.948200in, height=1.3422in,
width=4.16in ]{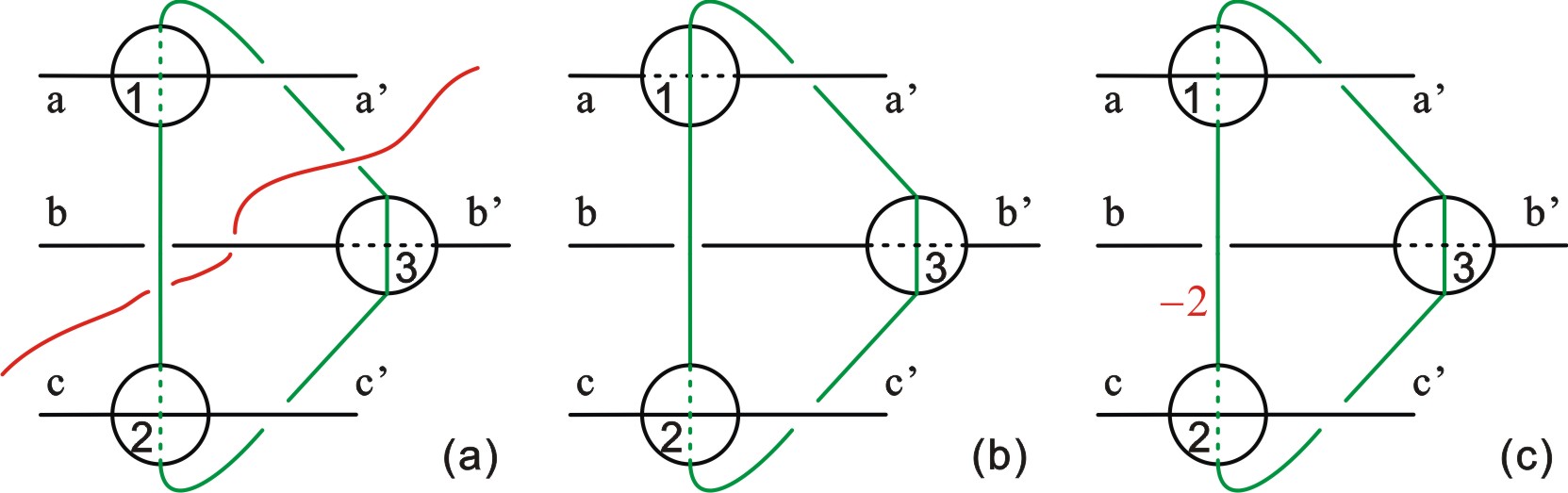}
\end{center}
\caption{(a) \& (b) are not configurations for a legal
$3\rightarrow2$ move due to respectively the red edge and edge $a'$
going throught the green loops, which makes the latter
incontractible; (c) is not either because of the $2\pi/3$-twist on a green edge.}%
\label{3to2not}%
\end{figure}

\item In the framed case there should be  no twist on the edges
forming the loop (such twists are henceforth called \textbf{internal
twists}); Fig. \ref{3to2not}(c) shows a counter-example.
\end{enumerate}
\end{condition}

\subsection{The $1\rightarrow4$ move}

\begin{figure}[h]
\begin{center}
\includegraphics[
natheight=1.522900in,
natwidth=2.948200in,
height=1.5584in,
width=2.9914in
]{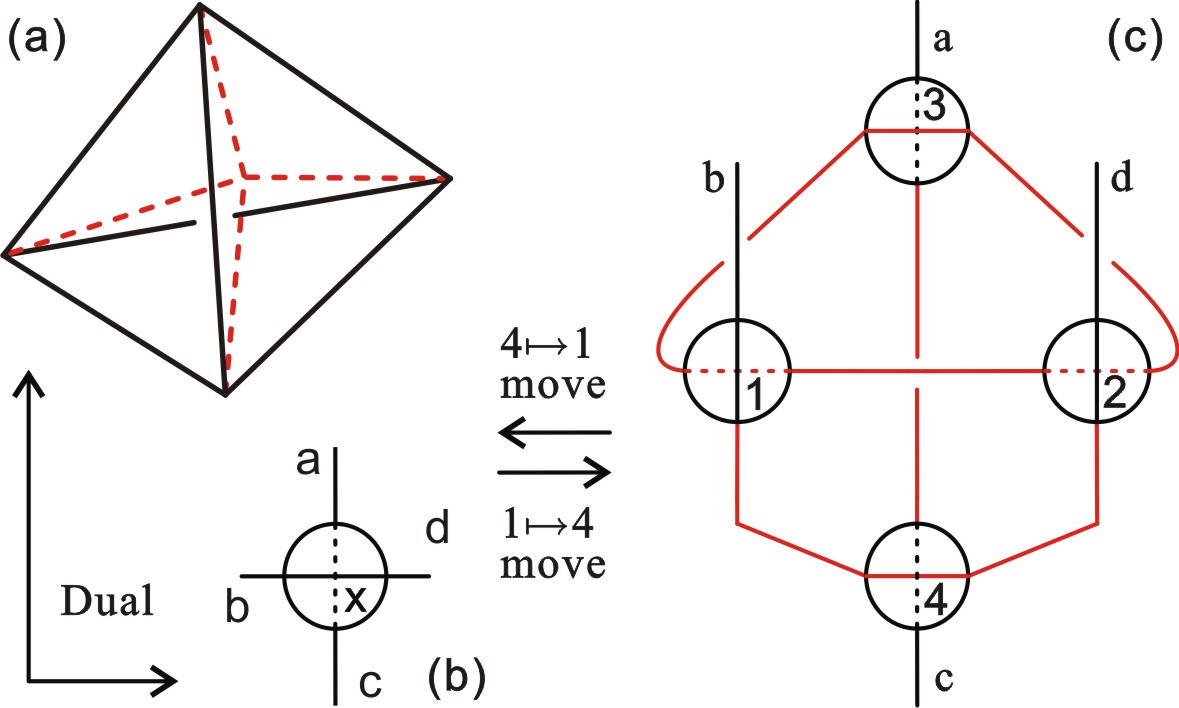}
\end{center}
\caption{(a) shows how a $1\rightarrow4$ move is viewed with tetrahedra;
dashed red lines illustrate the splitting of the big tetrahedron into four
tetrahedra. (b) is the node dual to the big tetrahedron in (a). (c) is
obtained from (b) by a $1\rightarrow4$ move, which can go back to (b) by a
$4\rightarrow1$ move. Red lines in (c) are the edges generated by the move.}%
\label{1to4+}%
\end{figure}

The $1\rightarrow4$ Pachner move is a decomposition of a single
tetrahendron into four tetrahedra contained in it. Its dual is a
move that takes a node and replaces it by four nodes, connected
together as a tetrahedra. This is illustrated in Fig. \ref{1to4+}.

The result of a $1\rightarrow4$ move is unique up to equivalence
moves. In Fig. \ref{1to4+}(c), one can find four loops formed by the
red edges, which are new edges generated by the move. Note that
there are no twists on the edges newly generated by the
$1\rightarrow4$ move. Note also that since a node is always taken to
be dual to a single tetrahedron, the basic rule is always satisfied
and there are no conditions constraining a $1 \rightarrow4$ move.

\subsection{The $4 \rightarrow1$ move}

The $4\rightarrow1$ Pachner move is the inverse of the
$1\rightarrow4$ move, and hence it requires finding an unusual
configuration of four tetrahedra each pair of which share a face.
There are only four faces left over, which make up a tetrahedron
which contains the original four. The dual $4\rightarrow1$ hence
also requires a special starting point, which is four nodes each
pair of which share an edge. As a result there are conditions
required for the $4\rightarrow1$ move, which are as follows.

\begin{condition}
\label{con4to1}A $4\rightarrow1$ move is doable on four nieghboring
nodes if and only if

\begin{enumerate}
\item the four nodes together with their common edges can be arranged as shown in Fig. \ref{1to4+}(c) or its parity inverse;

\item the loops are all contractible i.e. there is no any other edge going
through the loop or tangled with the loop; (See Fig. \ref{3to2not} for a
similar situation.)

\item there should be no internal twists on closed loops in the initial diagram.
\end{enumerate}
\end{condition}

\subsection{The unframed case and the role of internal twist}

The same evolution moves apply in the unframed case, since the
structure in the nodes is sufficient to define the dual Pachner
moves. One can also posit an extension of the class of theories just
described in which internal twist is allowed on closed loops that
are annihilated by the $3 \rightarrow 2$ move or $4 \rightarrow 1$
move.  We will see below examples in which this enlarges the class
of propagating braids.

\subsection{A conserved quantity}

Rotations create or annihilate twist and crossings simultaneously.
As a result, in the framed case, a certain combination of them is an invariant, which we
call the \textbf{effective twist number}\cite{yidun}. Given a connected subdiagram, $\cal R$, and an
arbitrary choice of orientations on the edges of $\cal R$, this is
defined by,
\begin{equation}
\Theta_{0}=\sum_{\substack{\text{all edges in
}\cal{R}}}T_{e}-2\times\sum_{\substack{\text{all Xings in }
\cal{R}}}X_{i},\label{theta0}%
\end{equation}
where $T_{e}$ is the twist number created by the rotation on an edge
of the sub-diagram, $X_{i}$ is the crossing number of a crossing
created by the rotation between any two edges in the sub-diagram,
and the factor of 2 comes from the fact that a crossing always
involve two edges. One can easily check that the rotations and
translations just defined, carried out within $\cal R$, preserve
$\Theta_{0}$.

$\Theta_{0}$ extends to a quantity which is conserved also under the
evolution moves, with the following modifications. Within a
subdiagram $\cal R$, the edges and nodes of the diagrams define
continuous curves where the edges continue across the diagrams as
noted (that is each node projection gives a choice of the four
incident edges into two pairs, each of which is connected through
the node.) These curves are of two kinds, open which end at two
points on the boundary of $\cal R$ and closed. In addition, define
an {\it isolated substructure}\cite{jonathan1} to be a subgraph
which is attached to the rest of the graph by a single edge.

\begin{itemize}

\item{}Choose an arbitrary orientation of the open curves.

\item{}Sum over the orientations of the closed curves.

\item{}In the countings of crossings all crossings of edges with edges in isolated substractures are to be ignored.

\end{itemize}
Then the following is conserved under all equivalence moves and
evolution moves that act entirely within $\cal R$.
\begin{equation}
\Theta_{\cal R}=\sum_{\text{Orientations of closed curves in }\cal
R} \left [ \sum_{\text{all edges in } \cal
R}T_{e}-2\times\sum_{\text{all Xings in } \cal R}X_{i} \right ]
\label{thetaR}
\end{equation}

\section{Stability of braids under the evolution moves}

Consider a braid of the form of Figure \ref{braid}. It is not difficult to show that
the subgraph formed by the two nodes and the three shared edges is not dual to
any triangulation of a ball in $R^{3}$. Hence a $2 \rightarrow3$ move cannot
be done on these two nodes. Nor can a $3 \rightarrow2$ move be done on any
three nodes that contain this pair, for the same reason. Thus, the braid is stable under single moves.

We can extend this to the general observation that any braid whose
graph is not dual to a triangulation of a ball in $\mathbb{R}^3$ is
stable under single allowed moves.  We believe that there is a
stronger stability result for such braids, but postpone
consideration of this problem to future work.

\section{Braid Propagation}

We now come to the results on propagation of braids that follow from
the evolution moves we have defined here. Because a braids can be
considered an insertion in an edge, it makes sense to speak of them
propagating to the left or to the right along that edge. To help
visualize this in the diagrams we will always arrange a braid so
that the edge of the graph it interrupts runs horizontally on the
page.

An important note must be made before we get into the details of
braid propagation. Recall that braid diagrams correspond to the same
diffeomorphism equivalence class if they are connected by a sequence
of equivalence moves. It is useful to pick out of each such
equivalence class a representative and we do so by requiring that
the braid diagram be joined to the larger graph by external edges
that are twist free; such a representative is unique for each
equivalence class of braids. This is motivated by the fact that the
dynamical moves that we will shortly see initiate propagation and
interactions require the external edges by which a braid connects to
the rest of the diagram are twist free. The statements that follow
concerning propagation and interaction refer to diagrams that
satisfy this external edge twist free condition.

We will see that the propagation is in some instances chiral, so
some braids propagate only to the left (right), while their mirror
images propagate only to the right (left).

\begin{figure}[ph]
\begin{center}
\includegraphics[
natheight=5.405100in,
natwidth=2.949900in,
height=5.4622in,
width=2.994in
]{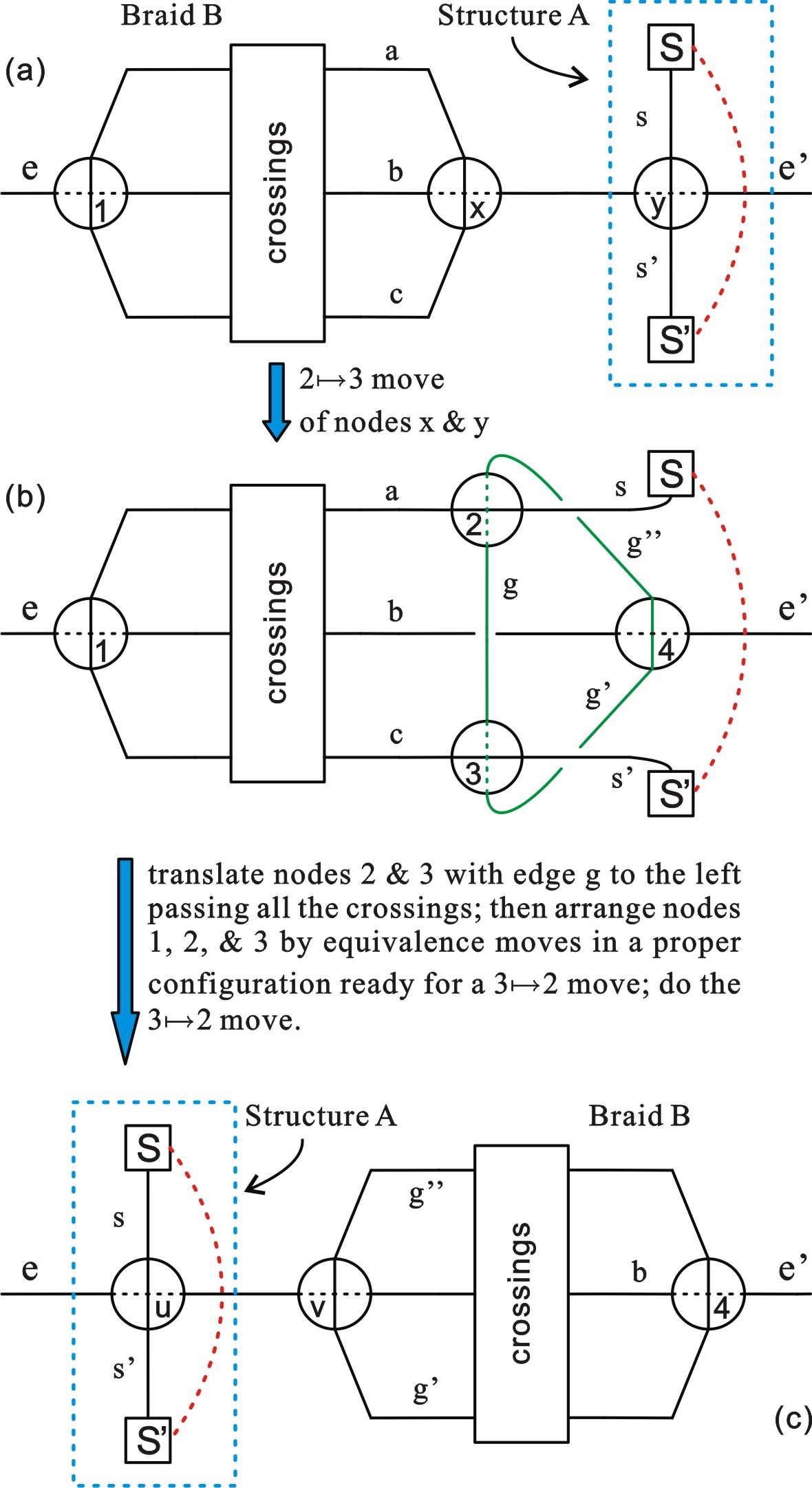}
\end{center}
\caption{This figure illustrates the process of a right-propagation. (a)
shows a braid $B$ sitting on the left of some arbitrary structure $A$,
composed of two sub-spinnets $S$ \& $S\prime$ connected via a node $Y$; the
right end-node of $B\,$, node $X$, and node $Y$ has a common edge; the dashed
red line means $S$ and $S^{\prime}$ may or may not be connected via another
edge. (b) is obtained from (a) by the $2\rightarrow3$ move on nodes $X$ and
$Y$. (c) is the result of the propagation, in which the braid $B$ and the
structure $A $ are both recovered and exchanged their relative positions, such
that $B$ is now on the right of $A$.}%
\label{propdef}%
\end{figure}

We study here a very simple example of propagation, which is  the
following. A braid is inserted in an edge of a graph as in Fig.
\ref{propdef}. We see that there is a node to the right of the
braid, with two structures growing out of it. By a series of local
moves, the braid moves so that these structures are now attached to
a node to the left of the braid, while the braid remains unchanged.

In more detail, a \textbf{right-propagation} of a braid $B$ involves
its changing places through a series of evolution moves with a
structure $A$ on its right, which is composed of two sub-spinnet $S$
and $S^{\prime}$connected by a node $Y$, as shown in Fig.
\ref{propdef}. The right end-node $X$ of the braid and node $Y$ have
one and only one common edge.

Begin with an initial condition as shown in Fig. \ref{propdef}(a).

\begin{figure}[h]
\begin{center}
\includegraphics[
natheight=1.452000in,
natwidth=2.952500in,
height=1.4875in,
width=2.9966in
]{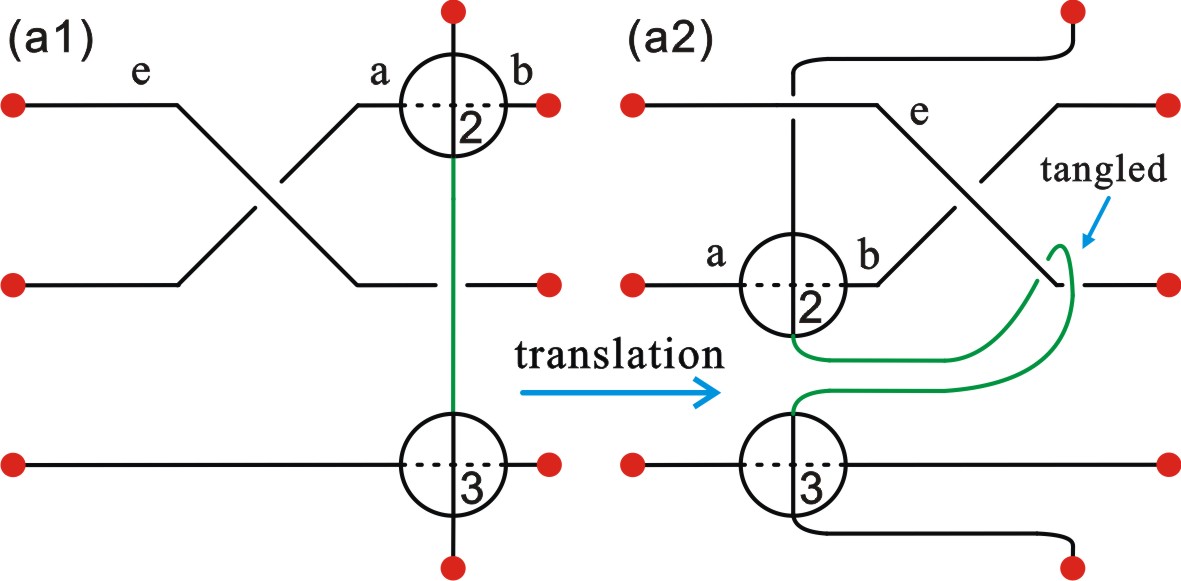}
\end{center}
\caption{(a2) is obtained from (a1) by translating nodes 2, 3, and their
common edge to the left, passing the crossing. A tangle occurs between the
edge $e$ and the common edge of nodes 2 and 3.}%
\label{translationfail}%
\end{figure}

\begin{enumerate}
\item Make a $2\rightarrow3$ move on nodes $X$ and $Y$ if the move is legal (See
Condition \ref{con2to3}. The result is shown in  Fig. \ref{propdef}(b).

\item Translate nodes 2 and 3 together with
their common edge $g$ to the left, passing all crossings. This may not be possible because  a
tangle between edge $g$ and any of the edges $a$, $b$, and $c$ occur (see
Fig. \ref{translationfail}).

\item Re-arrange nodes 1, 2, and 3 and their edges, by
equivalence moves (e.g. rotations), into a proper configuration
ready for a $3\rightarrow2$ move (see Condition \ref{con3to2}).

\item Do the $3\rightarrow2$ move on nodes 1, 2, and 3; leading to Fig.
\ref{propdef}(c), in which the braid $B$ and the structure $A$ are
both recovered but the braid $B$ is now on the right of structure
$A$.

\end{enumerate}

A braid that can propagate to its right is said to be
\textbf{right-propagating}.

Note that in Fig. \ref{propdef}(a), there is a dashed red line
connecting $S$ and $S^{\prime}$. This means that $S$ and
$S^{\prime}$ may possibly be directly connected or connected
throught another part of the whole spinnet. However, if $S$ and
$S^{\prime}$ are indeed so connected there are instances in which
the propagation of the braid cannot go through due to the tangle
between edges $s$, $s^{\prime}$ and the edges of the braid. Thus, if
not explicitly stated, the sub-spinnets $S$ and $S^{\prime}$ in the
structure $A$ are assumed to be only connected via the node $Y$.

A \textbf{left-propagation} is defined similarly in that the unknown
structure in a dashed blue square in Fig. (a) is intially on the
left of the braid.   A braid that can propagate to its left is then
called \textbf{left-propagatable}. A braid that can propagate to
both sides is a \textbf{two-way-propagatable} braid.

\subsection{Examples of chiral propagation}

We now give several examples of right-propagation.

\begin{figure}[h]
\begin{center}
\includegraphics[
natheight=2.987900in,
natwidth=2.955100in,
height=3.0312in,
width=2.9992in
]{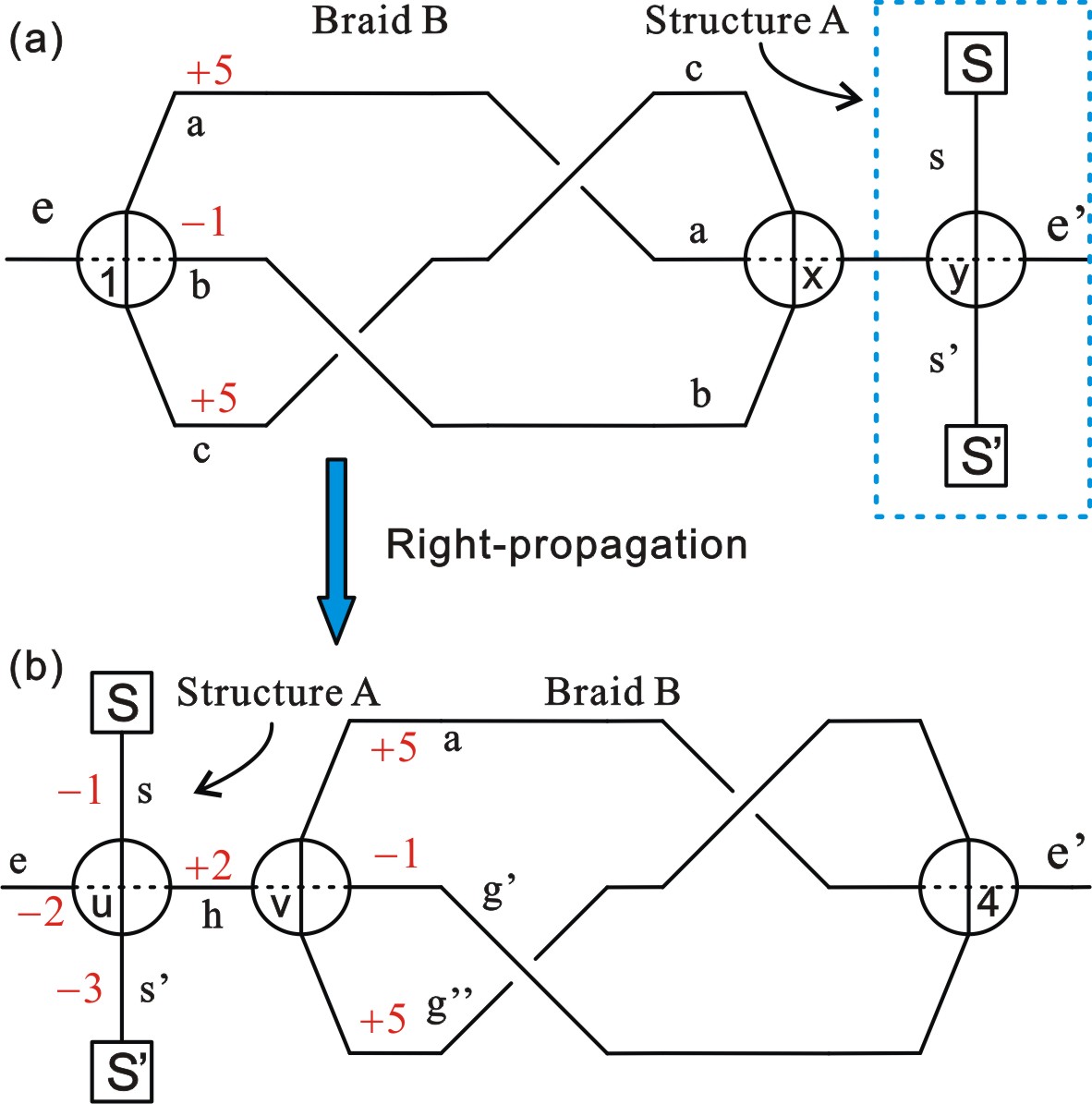}
\end{center}
\caption{A first example of braid propagation}%
\label{rightprop2xconserve}%
\end{figure}

\begin{figure}
[h]
\begin{center}
\includegraphics[
natheight=1.010100in, natwidth=4.602500in, height=1.0438in,
width=4.6553in
]%
{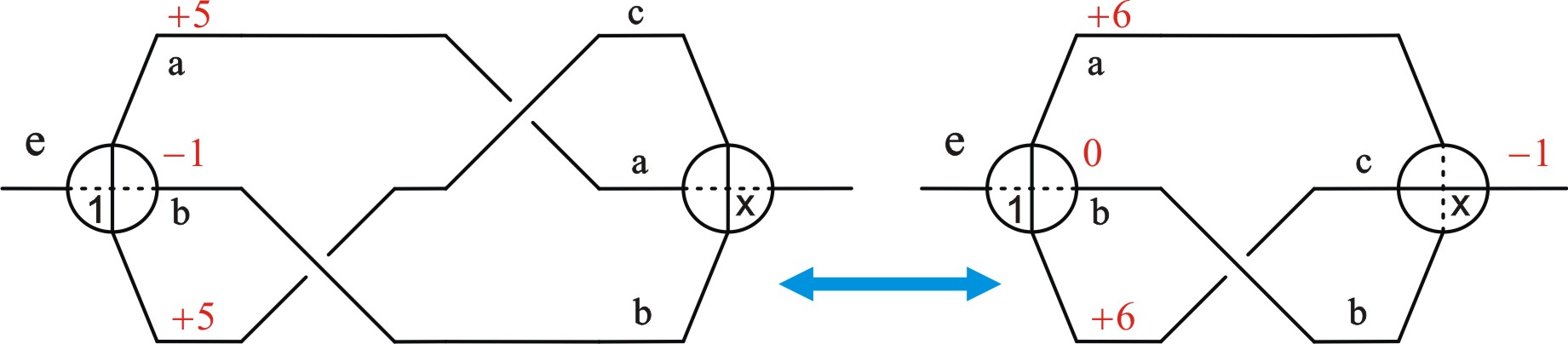}%
\end{center}
\caption{The equivalence between braid $B$ in Fig.
\ref{rightprop2xconserve},
shown on the left, and an irreducible 1-crossing braid, shown on the right.}%
\label{rightprop2xconserveequiv}%
\end{figure}

\begin{figure}
[h]
\begin{center}
\includegraphics[
natheight=3.137500in, natwidth=2.955100in, height=3.1825in,
width=2.9992in
]%
{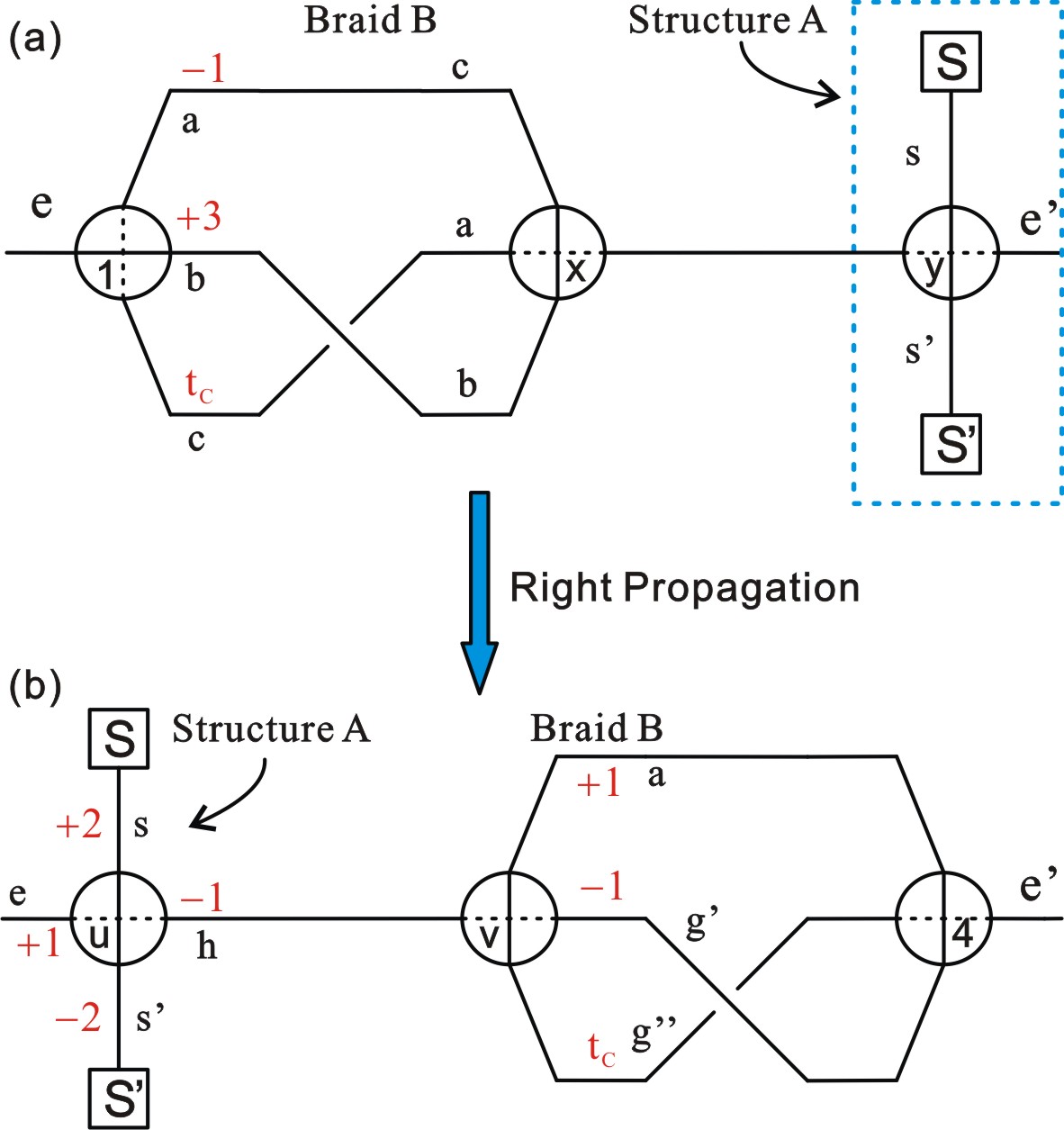}%
\caption{A second example of right propagation. Note that the twist
$t_{c}$ on
strand $c$ is arbitrary but the total twist is now not conserved.}%
\label{rightProp1XnoTconserve}%
\end{center}
\end{figure}

\begin{figure}[p]
\begin{center}
\includegraphics[
natheight=3.019900in,
natwidth=2.945600in,
height=8in,
width=5.84in
]{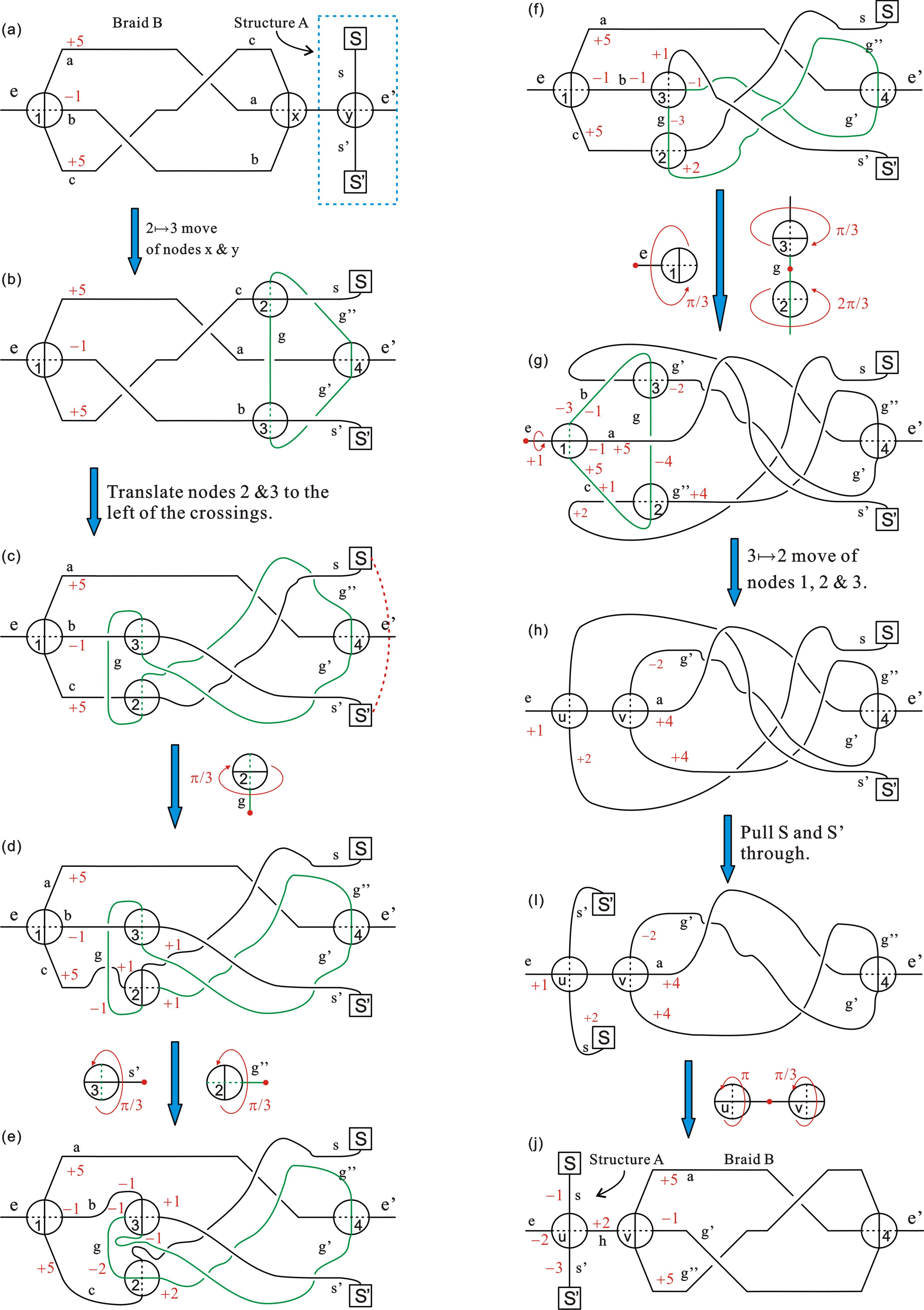}
\end{center}
\caption{The details of the first example of braid propagation, in
the framed case.} \label{rightprop2Xall}
\end{figure}

\begin{figure}[p]
\begin{center}
\includegraphics[
natheight=3.019900in,
natwidth=2.945600in,
height=8in,
width=5.84in
]{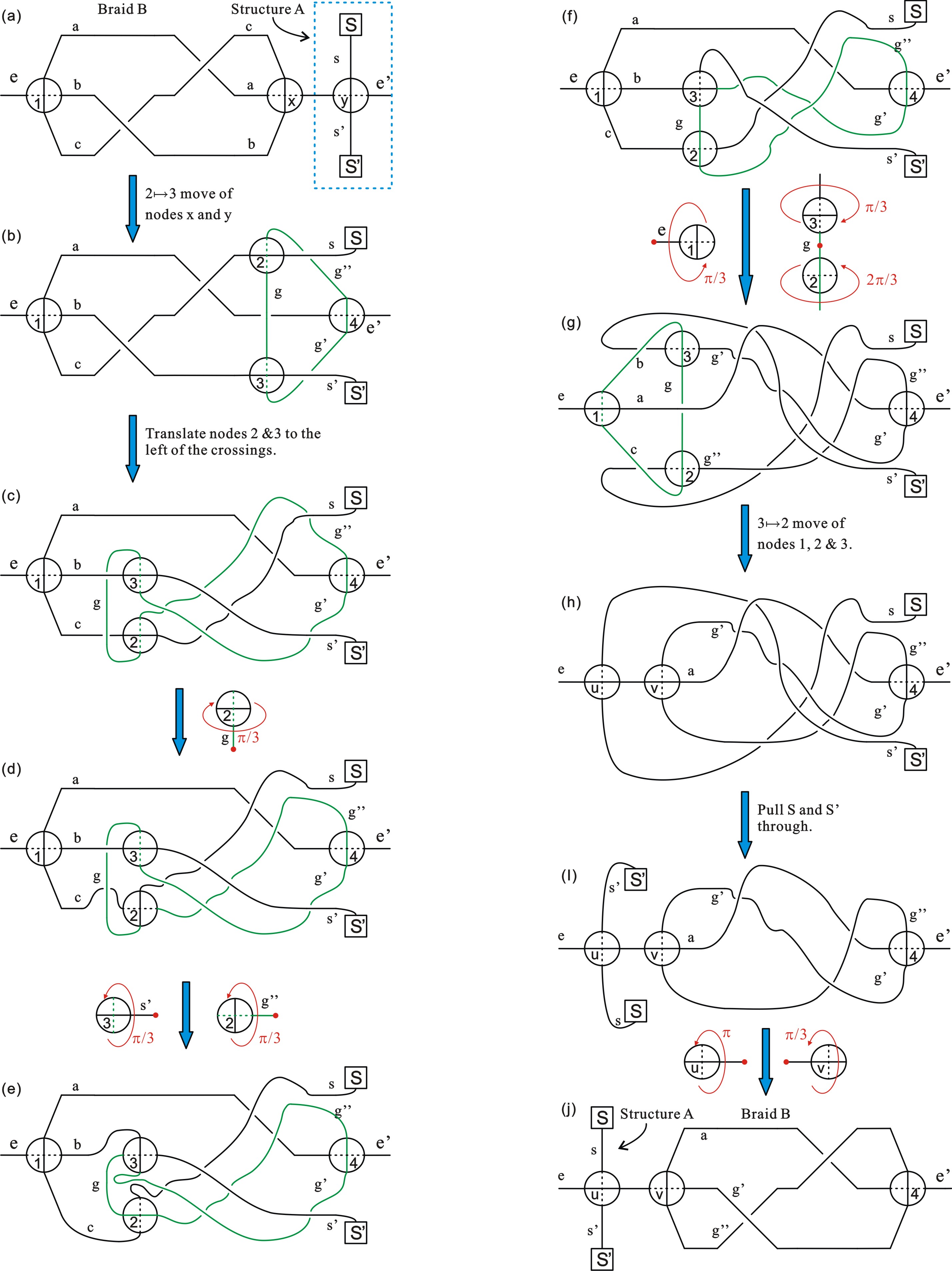}
\end{center}
\caption{The details of propagation in the unframed case.}
\label{rightprop2XnoTwistall}
\end{figure}

Our first example is illustrated in Fig. \ref{rightprop2xconserve}.
The evolution and equivalence moves required for the propagation of
shown in Fig. \ref{rightprop2xconserve} are given in Fig.
\ref{rightprop2Xall} for the framed case. For the unframed case they
are given in Fig. \ref{rightprop2XnoTwistall}. The braid $B$ in Fig.
\ref{rightprop2xconserve} propagates only to the right (the reason
will be explained by Theorem \ref{theoBraidProp} below).

In fact, it
is equivalent to an irreducible 1-crossing braid by a rotation on
its right end-node, as drawn in Fig. \ref{rightprop2xconserveequiv}.

In the framed case Fig. \ref{rightprop2xconserve} is one of the
examples in which internal twists are present, details of which are
illustrated in Fig. \ref{rightprop2Xall}(g). Hence the braid $B$ in
Fig. \ref{rightprop2xconserve}(a) will not propagate in the framed
case unless such moves are allowed. If we do allow the operation of
the $3\rightarrow2$ move in this situation, which leads to Fig.
\ref{rightprop2Xall}(h), the propagation succeeds and the initial
twists of braid $B$ are preserved individually, as shown in Fig.
\ref{rightprop2xconserve}. In the unframed case, such issues do not
arise so the braid is propagating.

A second example of right propagation is given in Fig.
\ref{rightProp1XnoTconserve}. This braid is in fact two-way
propagating, as can be easily checked. In this example, there is no
violation of Condition \ref{con3to2} due to nonzero internal twists.
Nevertheless, the initial twists of the propagating braid are not
individually preserved under the propagation.

\subsection{Which braids can propagate?}

In fact most braid diagrams propagate neither to the left nor to the
right, because of the following result.

\begin{theorem}
\label{theoBraidProp}A braid diagram, which is right-irreducible, in
a representation in which both external edges are untwisted, does
not  propagate to the right.
\end{theorem}
\begin{proof}
We prove this for the case of right-propagation; the left one
follows similarly. If a braid is not right-reducible, it means that
the braid has an irreducible 1-crossing right end-node. The idea is
to show that the crossing of each of these irreducible 1-crossing
end-nodes causes a tangle during translating the nodes generated by
a $2\rightarrow3$ move. There are four irreducible 1-crossing right
end-nodes in total, we pick one to demonstrate the proof, which is
depicted in Fig. \ref{profTheoIrredBraid}. In view of this, the
proof for the other three irreducible nodes are straightforward.
\begin{figure}[h]
\begin{center}
\includegraphics[
natheight=4.031800in, natwidth=2.945600in, height=4.0828in,
width=2.9888in ]{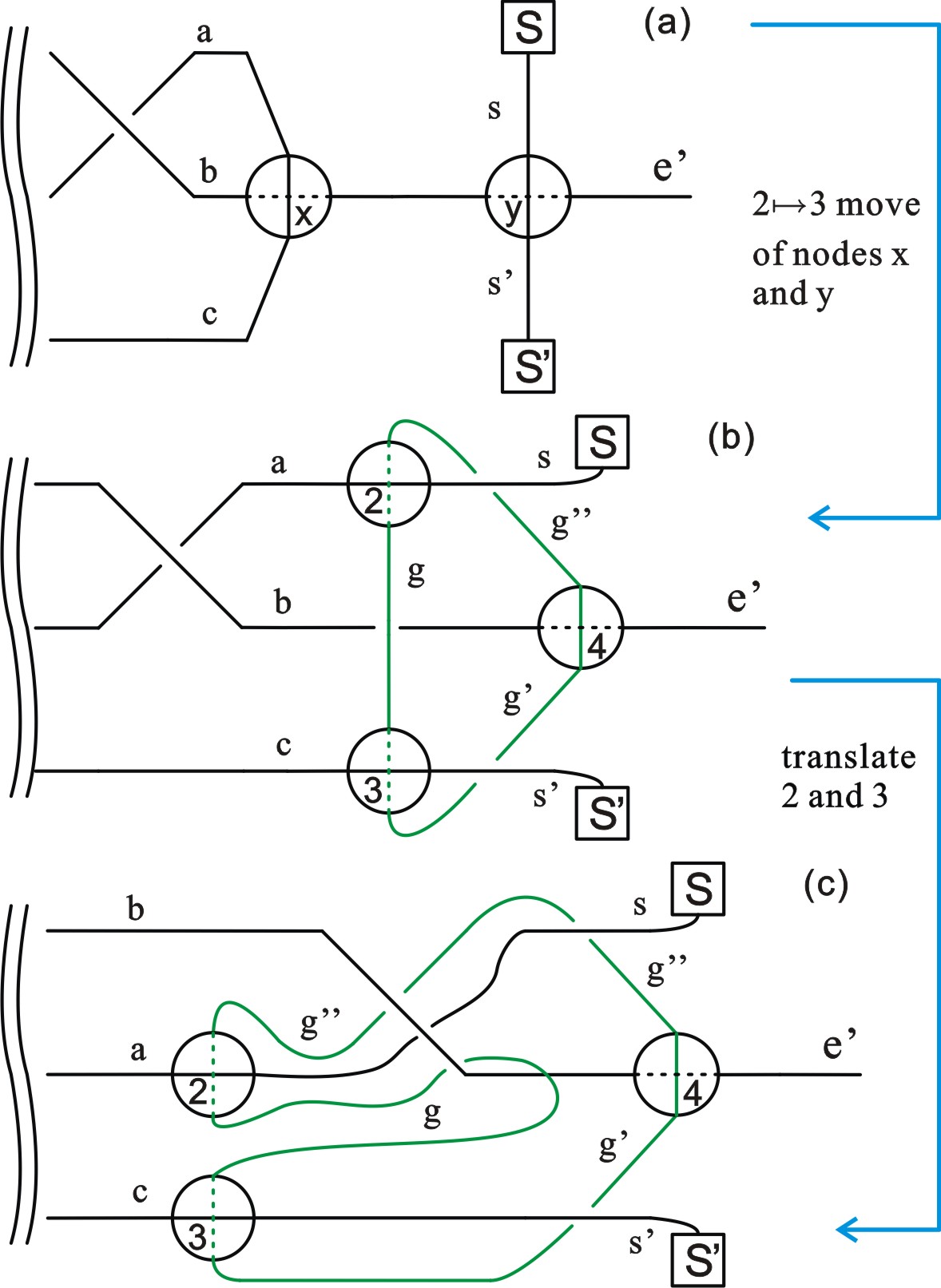}
\end{center}
\caption{(a) contains an irreducible 1-crossing right end-node,
which belongs to a braid whose left side is neglected, and a
structure as that in Fig. \ref{propdef}(a). (b) is obtained from (a)
by a $2\rightarrow3$ move on nodes $X$ and $Y$. (c) is the result of
translating nodes 2, 3 and edge $g $ to the left, passing the
crossing, which causes a tangle between edge $g$ and edge $b$. Note
that the crossing between edges $a$ and $b$ in Fig.
\ref{profTheoIrredBraid}(a) turns out to be a crossing between edges
$g^{\prime\prime}$ and $b$; a similar situation has been seen in
Fig. \ref{translationfail}.} \label{profTheoIrredBraid}
\end{figure}
\end{proof}

At first this result may seem puzzling. If a reducible braid diagram
$B$ propagates to the left, why doesn't the irreducible braid
diagram $B^\prime$ it reduces to?  Looking at Fig.
\ref{rightprop2xconserveequiv} we see that the point is that an
irreducible diagram can propagate, when it, together with its
environment, is equivalent to a diagram which has propagation. But
in those cases the requirement of having twist free external edges
is not met, precisely because the equivalence move required to take
the irreducible diagram to one that does propagate introduces a
twist.

This theorem has two immediate corollaries.

\begin{corollary}
\label{corIrredNotProp}An irreducible braid diagram, in a
representation in which both external edges are untwisted, does not
propagate either way.
\end{corollary}
\begin{proof}
An irreducible braid is neither left- nor right-reducible;
therefore, it does not propagate to either left or right, i.e. it
does not propagate.
\end{proof}

\begin{corollary}
\label{corRedProp}If a right-reducible braid diagram propagates, in
a representation in which both external edges are untwisted, it only
propagates to the right. The proof is obvious.
\end{corollary}

Examples of left-propagating braids are of course given by the
mirror images of the right-propagating braids. If one flips the
crossing on the left of braid $B$ in Fig.
\ref{rightprop2xconserve}(a), one obtain an example of two-way
propagating braids.

\bigskip

\section{Interactions}

\begin{figure}[h]
\begin{center}
\includegraphics[
natheight=4.954500in, natwidth=2.949900in, height=5.0099in,
width=2.994in ]{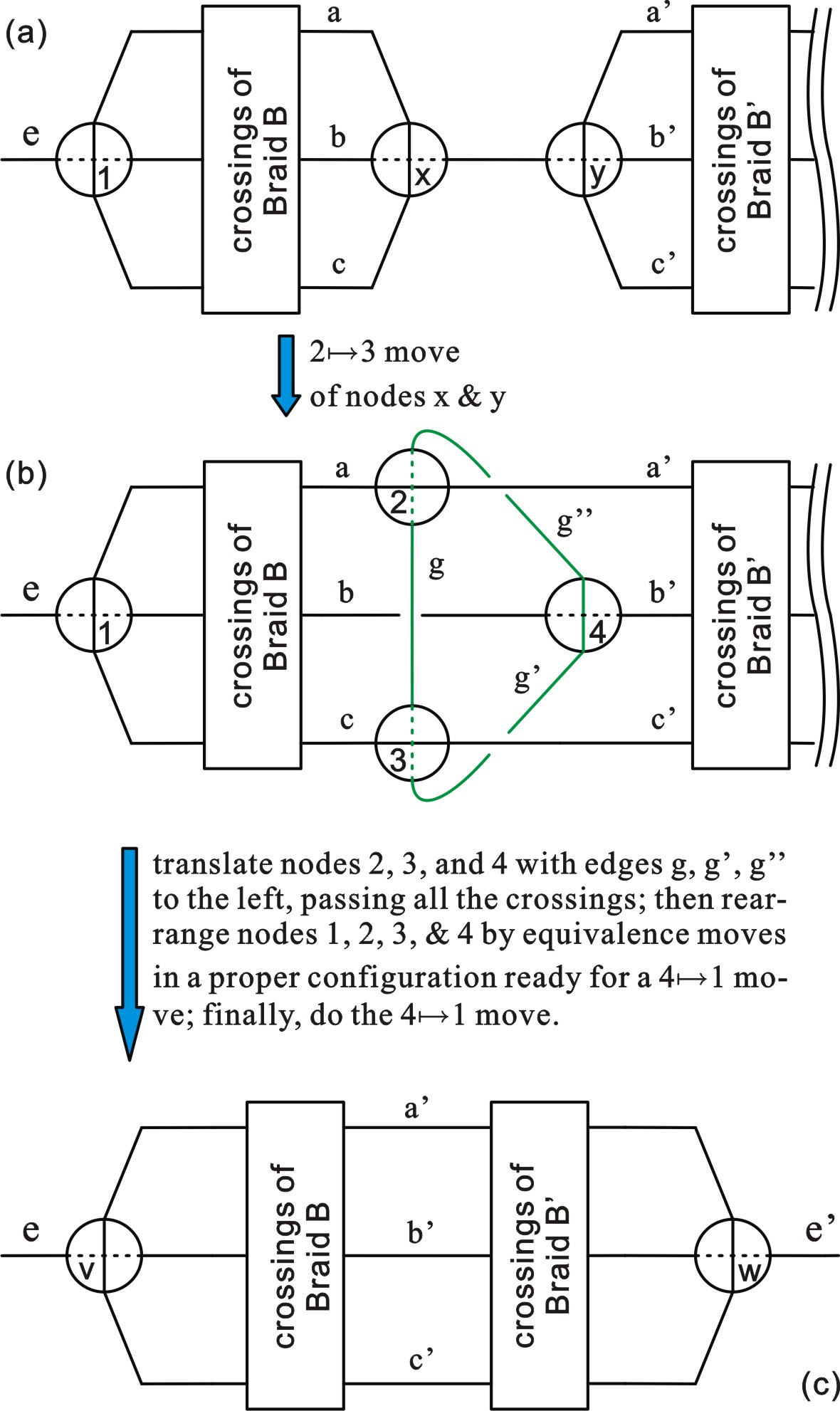}
\end{center}
\caption{The interaction of two braids.}
\label{intdef}
\end{figure}

We show now that in a few circumstances braids can interact with
each other, in the sense that two braids can meet and turn through a
sequence of local moves into a single braid. This will be possible
in the case that at least one of the braids are of a special class
called {\it actively interacting.}  As the moves are time reversal
invariant, it is also possible for a single braid to decay to two
braids, so long as one of the products is of this special class.

The initial condition for an interaction has two braids
adjacent to each other along an edge. Such a state can be reached by
either or both braids propagating along an edge. Then in some cases
there is a series of local moves resulting in a single braid. This
is illustrated in Figure \ref{intdef}. The time reversed history
will then give a case of a single braid decaying to two braids that
can then propagate separately.

It turns out that braids may interact actively or passively. If a
braid interacts actively, it can interact with any braid it
encounters, and hence are called {\it actively interacting.} All
other braids interact only passively, i.e. only if they encounter an
actively interacting braid. We will see shortly how and when a braid
can be actively interacting. We note also the possibility that
braids may interact from the left or right, so we distinguish those
two possibilities.

As in the case of propagation, the \textbf{active right-interaction}
of a braid $B$ with a neighboring braid $B^{\prime}$ proceeds
through a series of steps, indicated in Fig. \ref{intdef}. Each step
must be a possible evolution or equivalence move for the interaction
to take place.

\begin{enumerate}
\item Begin with the initial condition \ref{intdef}(a). As in the case of
propagation first make a $2\rightarrow3$ move on nodes $X$ and $Y$,
leading to the configuration shown in Fig. \ref{intdef}(b).

\item If possible, without creating any tangles of the kind shown in Fig. \ref{translationfail}, translate nodes 2, 3, and 4 together with
their common edge $g$, $g^{\prime}$, and $g^{\prime\prime}$ to the
left, passing all crossings of $B$.

\item If possible, re-arrange nodes 1, 2, 3, and 4 with their edges, by
equivalence moves into a configuration which allows $4\rightarrow1$
move.

\item If possible (see Condition \ref{con4to1}), do the $4\rightarrow1$ move on nodes 1, 2, 3, and 4,
resulting in Fig. \ref{intdef}(c).
\end{enumerate}

The actively interacting braid is the one which the three nodes are
pulled through, in this case it is braid $B$. We will see shortly
that the class of such braids is very limited. The other braid,
which may be arbitrary, is said to have passively interacted.

\subsection{Examples of interacting braids}

As we shall see in the next subsection, most braids do not actively
interact. Because the last step of an interaction is a $4
\rightarrow1$ move, the possibilities are limited because the
initial step of a $4 \rightarrow1$ move requires four nodes making
up a tetrahedron. Here the issue raised above of how restrictive are
the rules for a $4 \rightarrow1$ move comes in. If we are less
restrictive and allow internal twists, we get more actively
interacting braids.

Even allowing moves that annihilate internal twists, all the
examples we have so far found of actively interacting braids are
equivalent to the unbraid, made up of three strands with no
crossings, but with twists on the edges.

\begin{figure}
[h]
\begin{center}
\includegraphics[
natheight=2.757900in, natwidth=3.527600in, height=2.8003in,
width=3.5743in
]%
{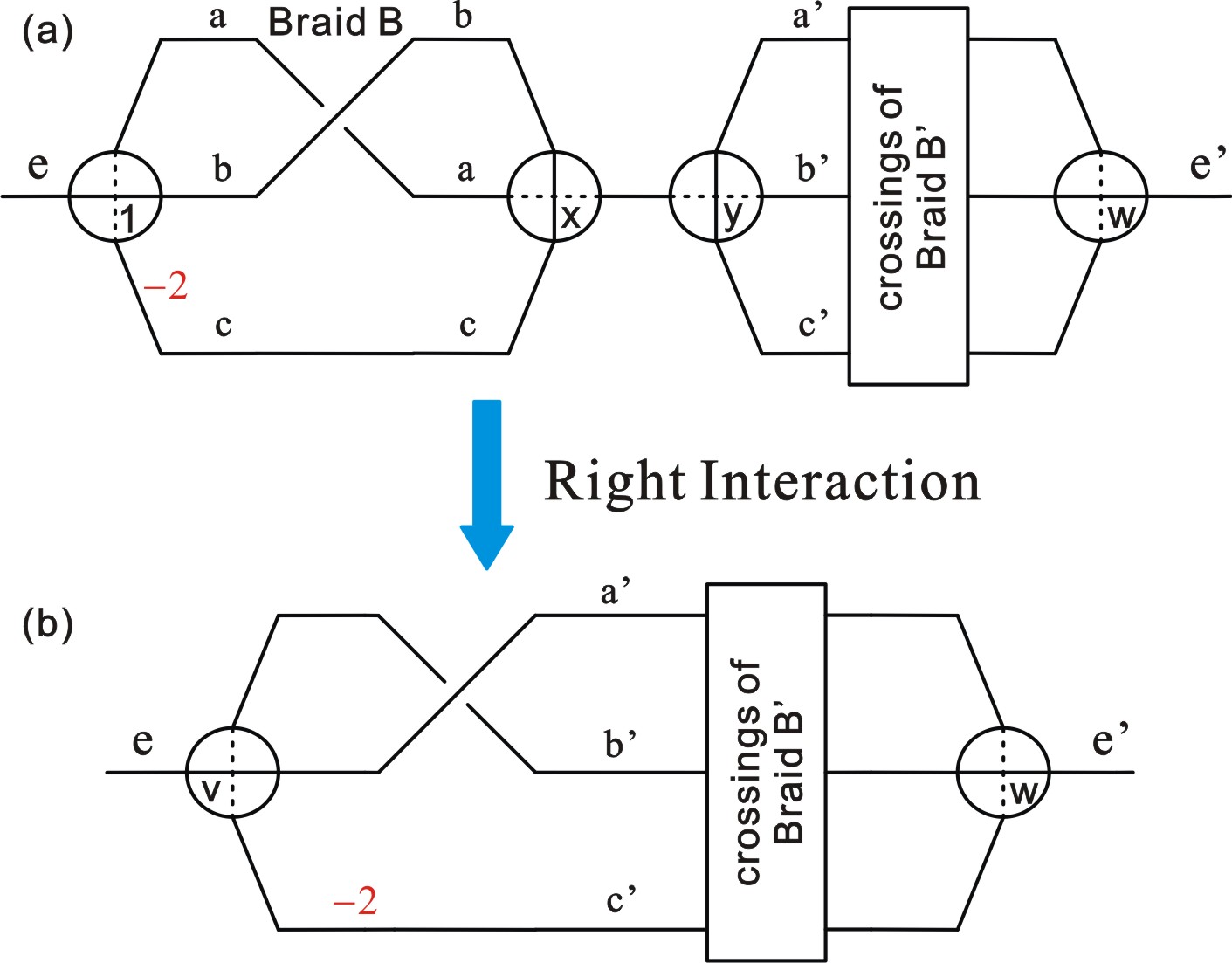}%
\caption{(a) contains a braid $B$ on the left and a braid
$B^{\prime}$ with arbitrary crossings \& twists on the right. (b) is
the resulted braid of the
active right-interatction of $B$ on $B^{\prime}$.}%
\label{rightInt1xLong}%
\end{center}
\end{figure}

Our first example of interaction is shown in Fig.
\ref{rightInt1xLong}. Fig. \ref{rightInt1xLong}(a) depicts two
braids: a 1-crossing braid $B$ on the left of a braid $B^{\prime}$
with arbitrary crossings and twists. The right end-node $W$ of braid
$B^{\prime}$ is irrelevant, which can be either $\oplus$ or
$\ominus$, here we assume it in the state $\ominus$ for the
convenience of showing the left-interaction later. One may notice
the $2\pi/3$-twist on edge $c$ of braid $B$, the presence of which
may be essential for the interaction to be done and will be
explained shortly. The active right-interaction of braid $B$ on
$B^{\prime}$ results in the new braid shown in Fig.
\ref{rightInt1xLong}(b); one can see that the initial twist of braid
$B$ in Fig. \ref{rightInt1xLong}(a) is preserved in Fig.
\ref{rightInt1xLong}(b)
under the interaction.%

\begin{figure}
[h]
\begin{center}
\includegraphics[
natheight=8.057500in, natwidth=7.727100in, height=5.246in,
width=5.0315in
]%
{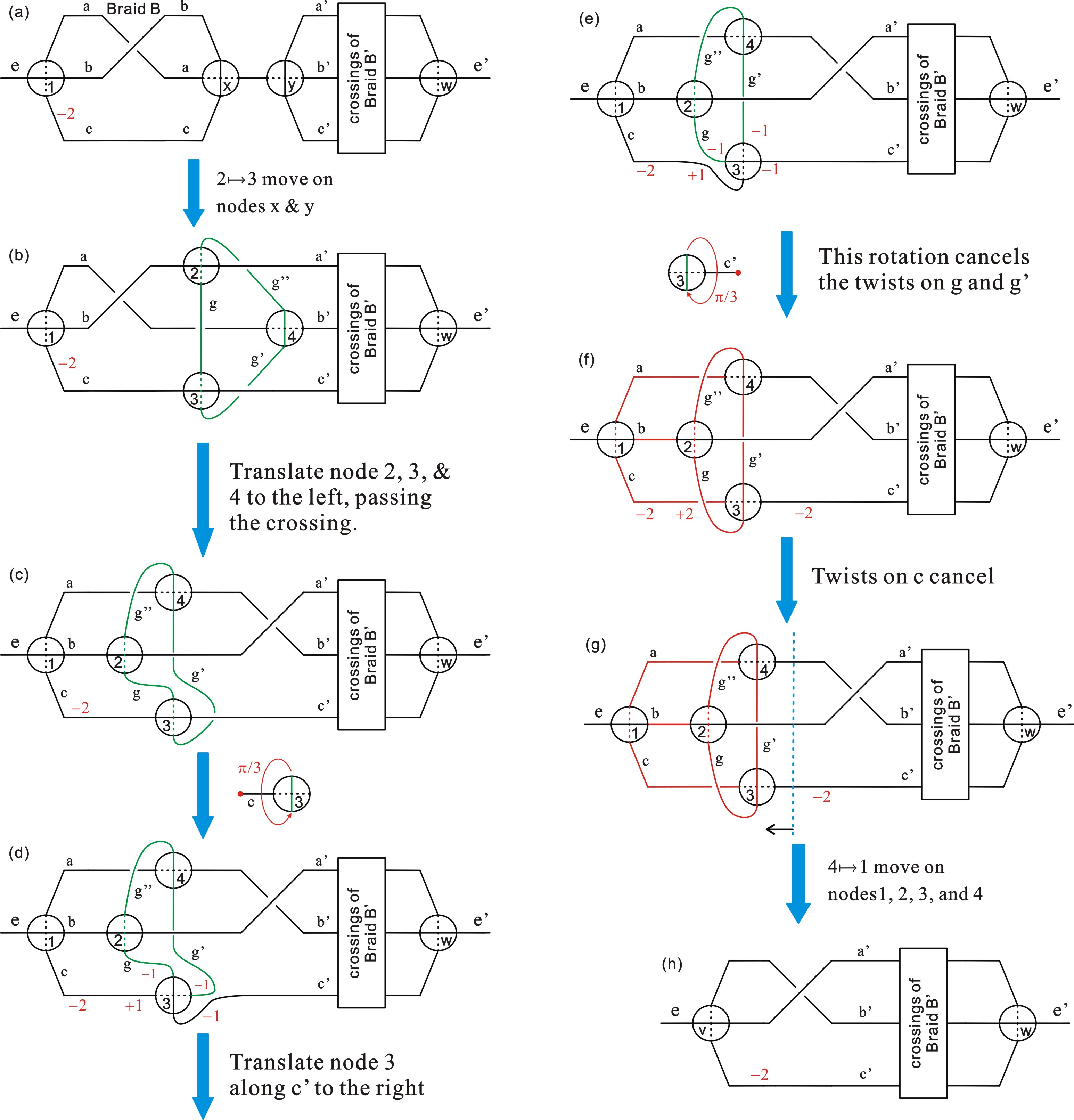}%
\caption{The details of the first example of interacting braids,
Fig.
\ref{rightInt1xLong}.}%
\label{rightInt1xLongAll}%
\end{center}
\end{figure}

The complete steps for the interaction are shown in Fig.
\ref{rightInt1xLongAll}. In Fig. \ref{rightInt1xLongAll}(f), the
initial twist on strand $c$, $-2$, of braid $B$ cancels the twist,
$+2$, created by rotations, such that nodes 1, 2, 3, and 4 with
their common edges are in a proper configuration for a
$4\rightarrow1$ move, satisfying Condition \ref{con4to1}. In
addition, surprisingly, as already mentioned, the initial twists of
braid $B$ are preserved.

\begin{figure}
[h]
\begin{center}
\includegraphics[
natheight=2.907500in, natwidth=3.527600in, height=2.9516in,
width=3.5743in
]%
{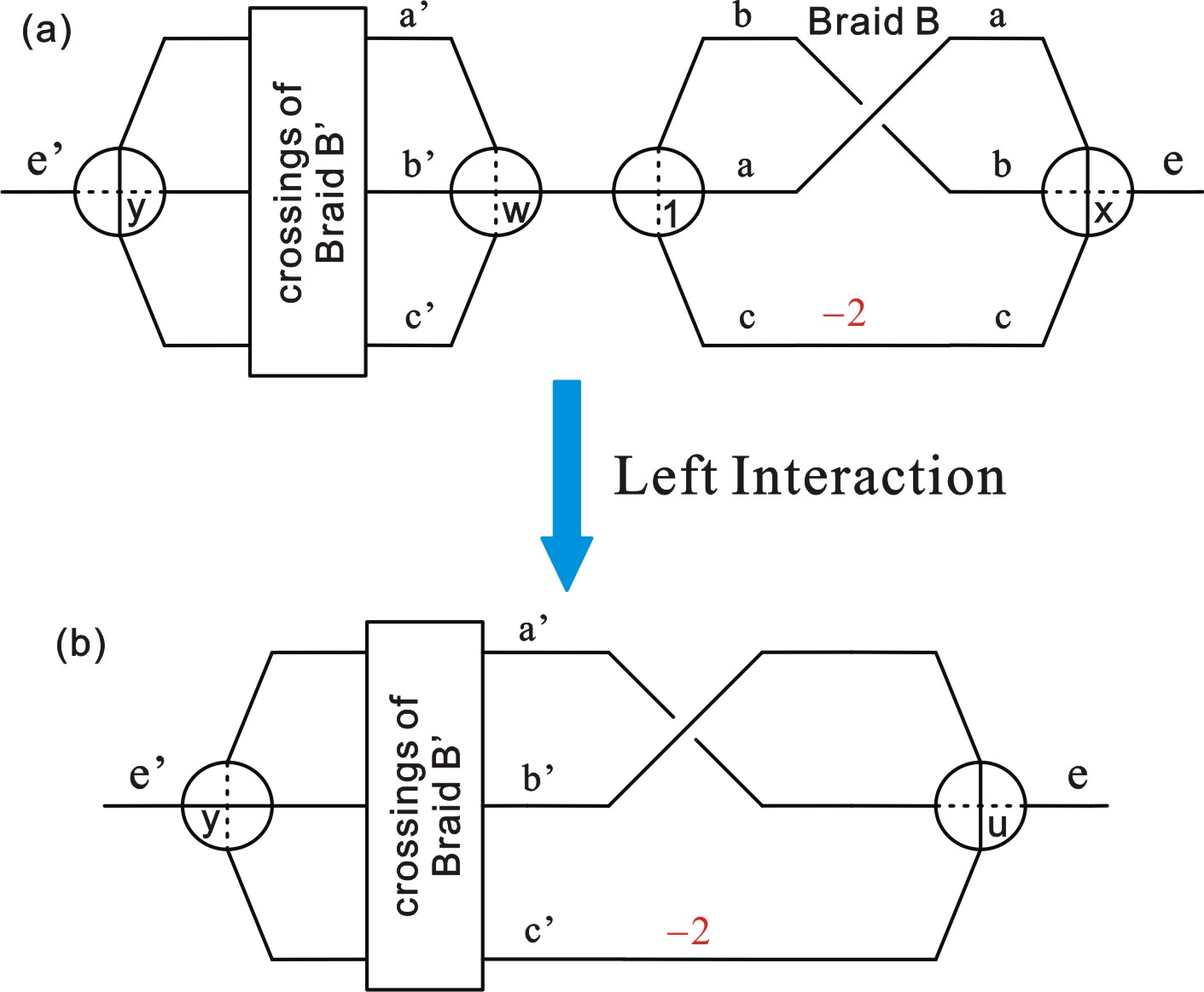}%
\caption{The braid in Fig. \ref{rightInt1xLong} can also interact from the left.}%
\label{leftInt1xLong}%
\end{center}
\end{figure}

The same braid can interact to the left as well, as we show in Fig. \ref{leftInt1xLong}.%

\begin{figure}
[h]
\begin{center}
\includegraphics[
natheight=0.927100in, natwidth=4.422700in, height=0.9599in,
width=4.4737in
]%
{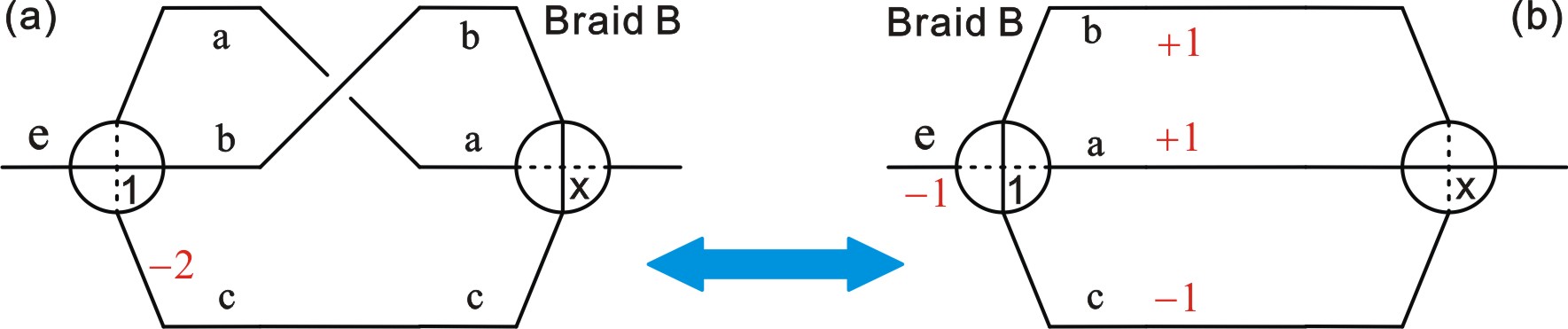}%
\caption{The braid in \ref{rightInt1xLong} is equivalent to a
trivial braid
with twists.}%
\label{rightint1xlongequiv}%
\end{center}
\end{figure}

Note that the braid $B$ is completely reducible from either end; if
we reduce it by a rotation of its right end-node, we get its
equivalent braid, Fig. \ref{rightint1xlongequiv}(b), which is a
trivial braid with twist numbers on all three strands and the right
external edge.

\begin{figure}
[h]
\begin{center}
\includegraphics[
natheight=2.804600in, natwidth=4.127700in, height=2.8478in,
width=4.1779in
]%
{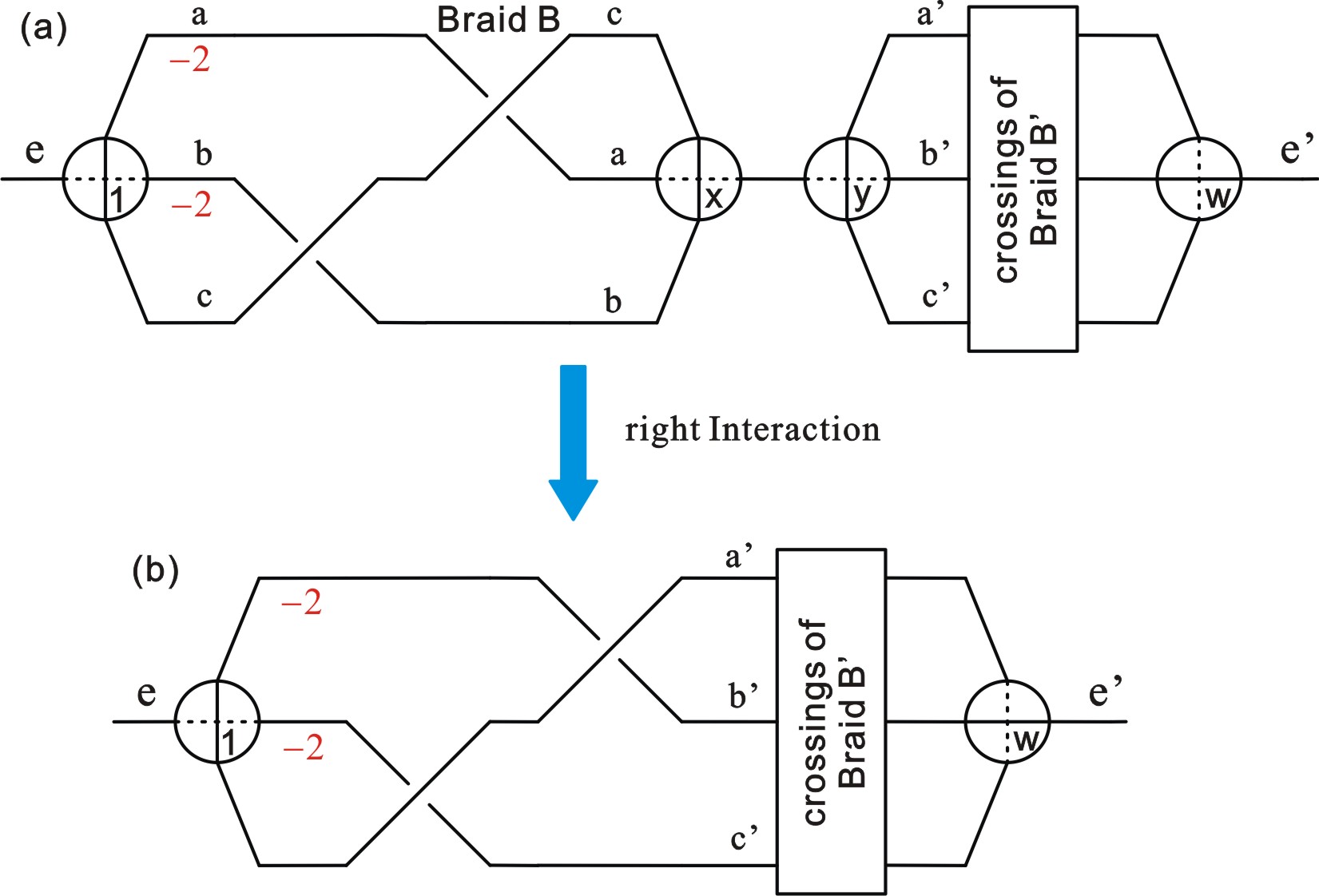}%
\caption{A second example of a right braid interaction.}%
\label{rightint2xlong}%
\end{center}
\end{figure}

A second example of right interaction is shown in Fig.
\ref{rightint2xlong}. This braid is also left-interacting, as can be
easily checked. And it is also equivalent to an unbraid with twists,
from top to bottom the twists are 0, 0, +2, respectively, and a
twist -2 on the right external edge. As for the first example, the
initial twists in this example cancel the possible internal twists
in the process of interaction, and they are individually conserved.

\subsection{General results on interactions}

Now we give some general results which limit the actively
interacting braids.

%\textbf{Corollary}
\begin{corollary}
\label{corIrredNotInt}A (left-) right-irreducible braid is not
(left-) actively right-interacting. An irreducible braid is always
inactive during an interaction. This also implies that only (left-)
right-reducible braids can be actively (left-) right-interacting,
which manifests the chirality of interactions.
\end{corollary}
\begin{proof}
This is a corollary of Theorem \ref{theoBraidProp}. As shown in Fig.
\ref{intdef}, to complete a (left-) right-interaction requires the
translation of nodes 2, 3, and 4 together with their common edges
$g$, $g^{\prime}$, and $g^{\prime\prime}$. On the other hand, in the
proof of Theorem \ref{profTheoIrredBraid} we showed that as long as
a braid is (left-) right-irreducible, a tangle between edge $g$ and
original edges of braid $B$. Consequently, the translation towards
an right-interaction of braid $B$ will at least cause the tangle
between edge $g$ and strands of the braid, which prevents the final
$4\rightarrow1$ move according to Condition \ref{con4to1}; hence,
the right-interaction cannot be completed as desired. It is the same
for a left-interaction. Therefore, is a braid is irreducible, i.e.
both left- and right-irreducible, the braid can never actively
interact onto anther braid; it always behaves inactively in any type
of interactions.
\end{proof}

%\textbf{Theorem 3:}
\begin{theorem}
\label{theoInt2Prop}If a braid is actively (left-/right-)
two-way-interacting, it is also (left-/right-) two-way-propagating.
\end{theorem}
\begin{proof}
The proof contains two parts.

1) We notice that in doing an interaction one needs to translate all
the three nodes (say nodes $\alpha$, $\beta$, and $\gamma$) and
their common edges, obtained from the initial $2\rightarrow3$ move,
all the way through the crossings of the active braid, say braid
$B$. However, if braid $B$ is to propagate in the same direction,
two of the nodes $\alpha$, $\beta$, and $\gamma$ must be
successfully translated. Therefore, if braid $B$ is able to actively
interact to the left (right), i.e. the translation of $\alpha$,
$\beta$, and $\gamma$ does not cause any tanglement, the translation
of any two of the three nodes should not lead to any tanglement
either.

2) Given part 1), if\ braid $B$ can actively interact, after being
translated nodes $\alpha$, $\beta$, and $\gamma$ together with an
end-node (say node 1) of braid $B$ can be arranged in a proper
configration for a $4\rightarrow1$ move. There is no loss of
generality to assume that we need to translate $\alpha$ and $\beta$
for $B$ to propagate in the same direction as its interaction;
hence, after the translation we must demand that $\alpha$, $\beta$,
and node 1 can be rearranged in a proper configuration for a
$3\rightarrow2$ move. So the question is, if $\alpha$, $\beta$,
$\gamma$ and 1 can be configured for a $4\rightarrow1$ move, can we
also configure $\alpha$, $\beta$, and 1 for a $3\rightarrow2$ move?

Let us consider active right-interaction. In the proper
configuration for a $4\rightarrow1$ move, we rename $\alpha$,
$\beta$, and $\gamma$ by the permutation of 2, 3 and 4, since we
don't know the relative positions of $\alpha$, $\beta$, and $\gamma$
after their translation over the crossings of an arbitrary braid
$B$; the situation is shown in Fig. \ref{profInt2Prop1}(a). We
assumed the propagation of braid $B$ requires translating $\alpha$
and $\beta$; however, we don't know really know the renaming of
$\alpha$ and $\beta$ after their translation within an abitrary
braid $B$; they can be 2 and 3, or 2 and 4, or 3 and 4. Therefore,
to prove that right-interaction implies right-propagation, it
suffices to show that one can always re-arrange nodes 1, 2 and 3, or
1, 2, and 4, or 1, 3, and 4 in a proper configuration for a
$3\rightarrow2$ move if 1, 2, 3 and 4 can be configured as in Fig.
\ref{profInt2Prop1}(a).

Fig. \ref{profInt2Prop1} cleary illustrates the procedure of
re-arranging nodes 1, 3, and 4 in a proper configuration for a
$3\rightarrow2$ move (compare the green loop in (c) with Fig.
\ref{2to3--}(b)), from the proper configuration of 1, 2, 3, and 4
for a $4\rightarrow1$ move. Interestingly and fortunately, all the
equivalence moves taking 1, 3, and 4 in \ref{profInt2Prop1}(a) to
them in (c) do not result in any twist along the green loop, which
satisfies Condition \ref{con3to2}; this is a subtlety of the proof.
As a consequence of symmetry, nodes 1, 2, and 4 in Fig.
\ref{profInt2Prop1}(a) can also be configured properly for a
$3\rightarrow2$ move. The case of re-arranging nodes 1, 2, and 3 in
Fig. \ref{profInt2Prop1}(a) in a proper configuration for a
$3\rightarrow2$ move is depicted in Fig. \ref{profInt2Prop2}.

The proof in the case of an active left-interaction follows
immediately by similarity and symmetry. Therefore, the theorem is
proved.

As a closing remark, the validity of the theorem can also be seen if
one looks at the local dual picture of a $4\rightarrow1$ move, in
which four tetrahedra stick together forming a new tetrahedron; any
three of them can form two tetrahedra by a $3\rightarrow2$ Pachner
move. The above proof simply interpret this topological picture in a
clear way with our notation of the embedded spinnets. In fact, the
examples we have found of actively interacting braids are all
reducible to an unbraid with twists. This is consistent with these
general results.
\begin{figure}[h]
\begin{center}
\includegraphics[
natheight=2.903200in, natwidth=4.775500in, height=2.9464in,
width=4.8291in ]{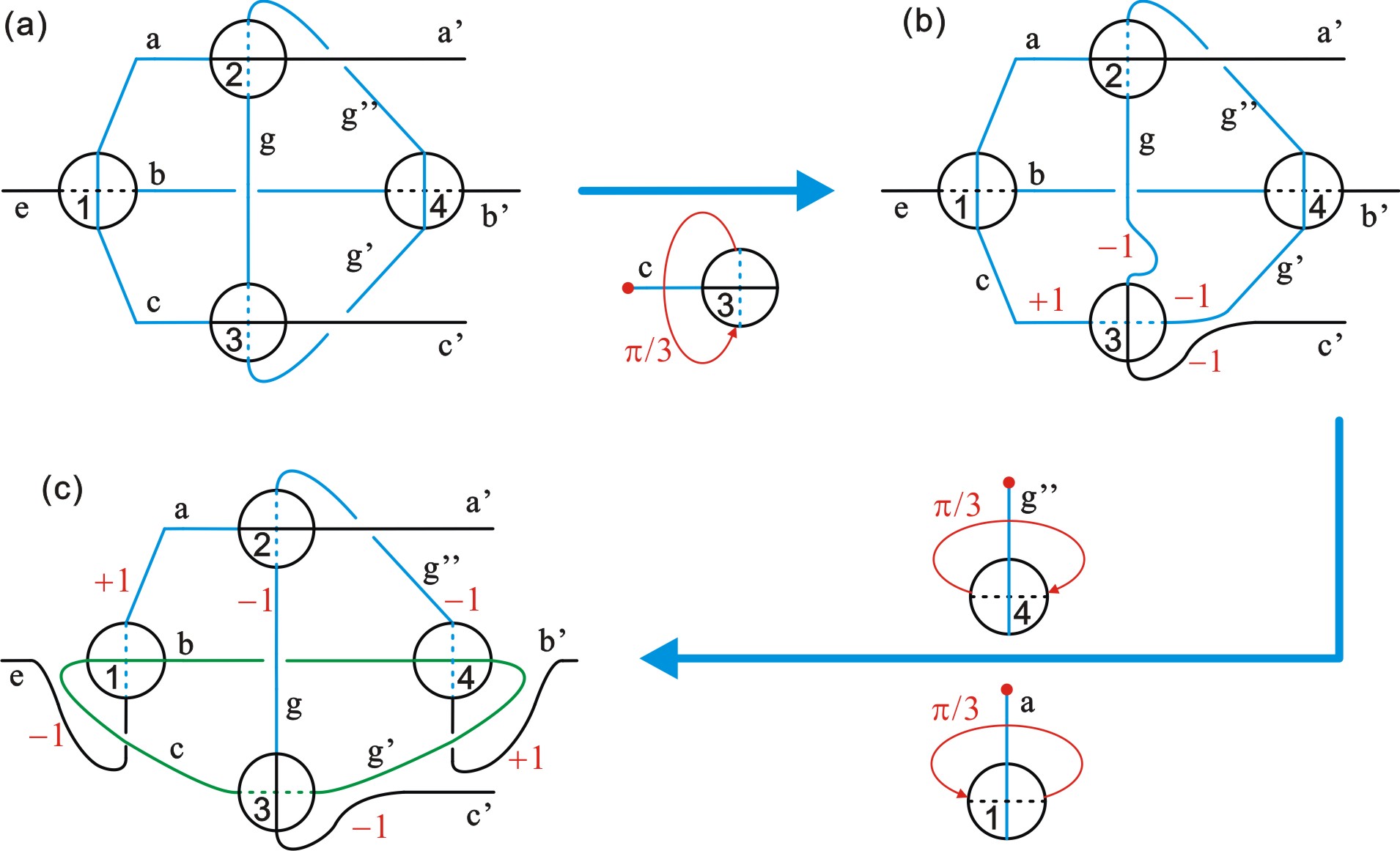}
\end{center}
\caption{This figure shows the full procedure of re-arranging nodes
1, 3, and 4 in (a) in a proper configuration for a $3\rightarrow2$
move, which is illustrated in (c). Note that the green loop in (c)
is twist free after
rotating nodes 1 and 4 from (b).}%
\label{profInt2Prop1}%
\end{figure}\begin{figure}[h]
\begin{center}
\includegraphics[
natheight=3.025100in, natwidth=4.770300in, height=3.0692in,
width=4.8239in ]{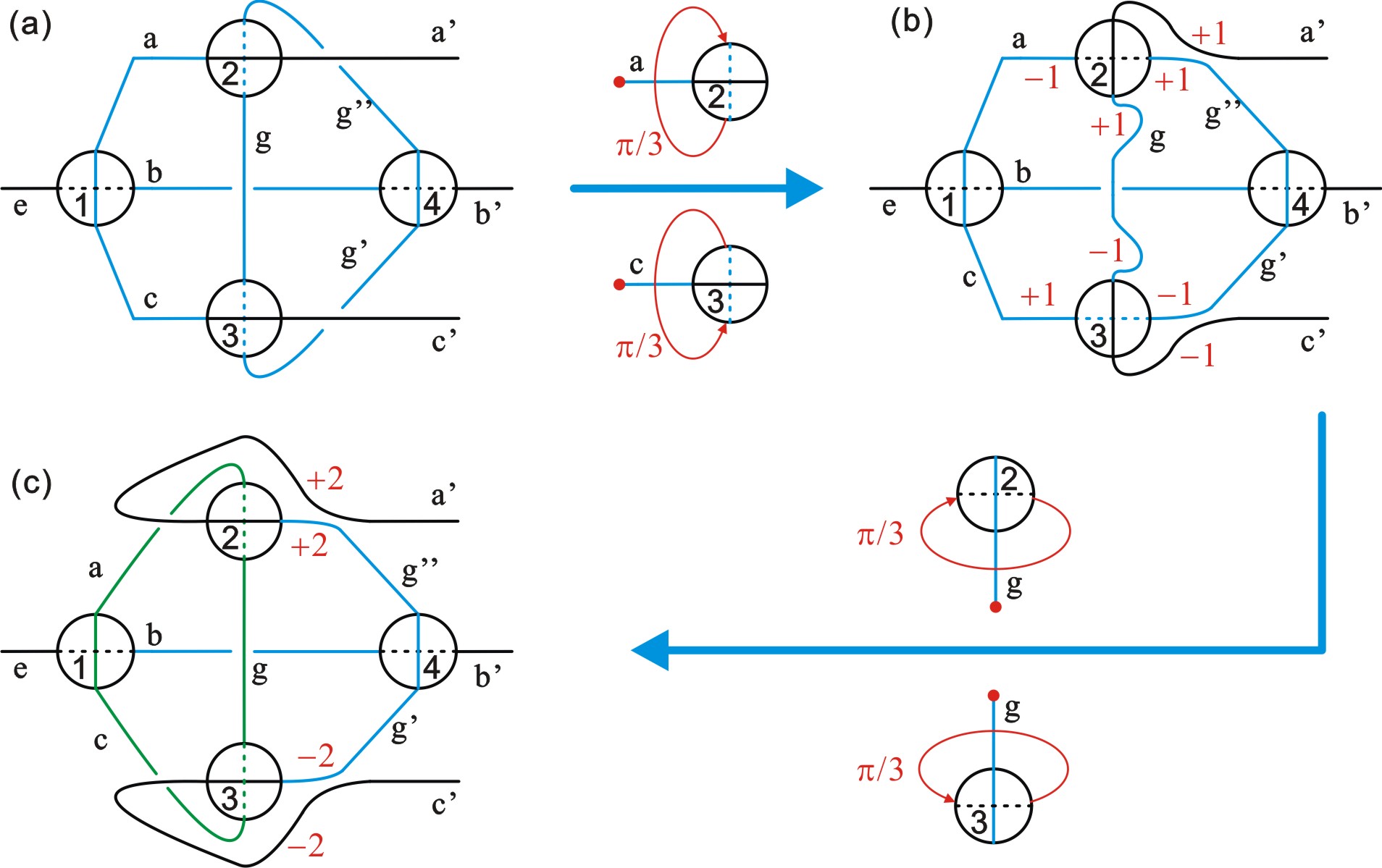}
\end{center}
\caption{This figure shows the full procedure of re-arranging nodes
1, 2, and 3 in (a) in a proper configuration for a $3\rightarrow2$
move, which is illustrated in (c). Note that the green loop in (c)
is twist free after
rotating nodes 2 and 3 from (b).}%
\label{profInt2Prop2}%
\end{figure}
\end{proof}

\section{Conclusions}

We have considered here the interaction and propagation of locally
topologically stable braids in four valent graphs of the kind that
occur in loop quantum gravity and spin foam models. The dynamics is
restricted to those coming from the dual Pachner moves, which
includes  the widely studied spin foam models. The graphs
representing the states are embedded in a three manifold up to
diffeomorphisms. We considered both the framed and unframed cases.
We also neglected the labels or colours on the spin networks as our
results do not depend on them, nor do they depend on the precise
amplitudes of the dual Pachner moves, so long as they are
non-vanishing.

We studied braids made up of two four valent nodes sharing three edges and found there are three classes of graphs.

\begin{itemize}

\item{} Actively interacting braids, which also propagate.  All the examples found so far are completely reducible to an unbraid with twisted edges. This means that they are characterized by three integers which give the twists on the three edges.

\item{}Braids which propagate but are not actively interacting.  The possibilities for these are limited by general results we found.

\item{}Braids, which are neither propagating or interacting.

\end{itemize}

It is very interesting to note that the braids required in the
three-valent case to realize Bilson-Thompson's preon
mode\cite{sundance} also are classified by three integers
representing twists on an unbraid\cite{JLS-inprogress}. This
suggests that it may be possible to incorporate a preon model with
interactions within the dynamics of four-valent braids studied here.

There are other interesting research lines in this direction. For
example, in usual settings of spinfoam models, tensor models or
group field theories, graphs are unembedded and labeled by elements
of the appropriate permutation groups, and are then called fat
graphs. There is an explicit criterion to decide whether the
simplicial complex associated to a given fat graph as a Feynman
diagram of a suitable field theory is a manifold or
not\cite{depietri}. It would be very interesting to clarify the
relationship between the unembedded fat graphs and our embedded
(framed) graphs and to see if our results also apply in the usual
spinfoam or group field theory settings. Recently, Premont-Schwarz
has also obtained related results in the unembedded
case\cite{isabeau}.

\section*{Acknowledgements}

We thank Fotini Markopoulou for initial collaboration on this
project, including the suggestion to examine braids in four valent
graphs with a restricted set of dual Pachner moves. We are also
grateful for discussions and comments with Sundance Bilson-Thompson,
Jonathan Hackett, Louis Kauffman and Isabeau Premont-Schwarz.
Research at Perimeter Institute is supported in part by the
Government of Canada through NSERC and by the Province of Ontario
through MEDT.

\end{document}